\documentclass[aps,superscriptaddress,showpacs,nofootinbib,eqsecnum]{revtex4}

\usepackage{amssymb}       
\usepackage{epsf,epsfig}      
\usepackage{graphics}
\def\case#1/#2{{\textstyle {#1\over #2}}}

\begin{document}

\title{Warm Inflation and its Microphysical Basis}

\author{Arjun Berera} \email{ab@ph.ed.ac.uk} \affiliation{School of Physics and Astronomy,
  University of Edinburgh, Edinburgh, EH9 3JZ, United Kingdom}

\author{Ian G.  Moss} \email{ian.moss@ncl.ac.uk} \affiliation{School of
  Mathematics and Statistics, University of Newcastle Upon Tyne, NE1 7RU,
  United Kingdom}

\author{Rudnei O. Ramos} \email{rudnei@uerj.br} \affiliation{Departamento de
  F\'{\i}sica Te\'orica, Universidade do Estado do Rio de Janeiro, 20550-013
  Rio de Janeiro, RJ, Brazil}

\begin{abstract}
  The microscopic quantum field theory origins of warm inflation dynamics are
  reviewed.  The warm inflation scenario is first described along with its
  results, predictions and comparison with the standard cold inflation
  scenario.  The basics of thermal field theory required in the study of warm
  inflation are discussed.  Quantum field theory real time calculations at
  finite temperature are then presented and the derivation of dissipation and
  stochastic fluctuations are shown from a general perspective.  Specific
  results are given of dissipation coefficients for a variety of quantum field
  theory interaction structures relevant to warm inflation, in a form that can
  readily be used by model builders.  Different particle physics models
  realising warm inflation are presented along with their observational
  predictions.
\end{abstract}


\maketitle

\vspace{0.1cm}

keywords: cosmology, inflation, dissipation, particle production, supersymmetry

\medskip

In Press Reports on Progress in Physics, 2009

\section{Introduction}

It has been fourteen years since warm inflation was introduced and with it the
first and still only alternative dynamical realisation of inflation to the
standard scenario.  The standard picture of inflation introduced in 1981
relied on a scalar field, called the inflaton, which during inflation was
assumed to have no interaction with any other fields.  During inflation, this
field rolls down its potential and due to it being coupled to the background
metric, a damping-like term is present which slows down its motion.  As this
inflaton field was assumed to not interact with other fields, there was no
possibility for radiation to be produced during inflation, thus leading to a
thermodynamically supercooled phase of the Universe during inflation.  Getting
out of this inflation phase and putting the Universe into a radiation
dominated phase was a key issue, termed the ``graceful exit" problem
\cite{oldinf,ni,reheatu}.  The first successful solution of the graceful exit
problem and so the first successful cold inflation model, was new inflation
\cite{ni}.  The solution was to picture particle production as a distinct
separate stage after inflationary expansion in a period called reheating.  In
this phase, couplings to other fields were assumed to be present and the
inflaton would find itself in a very steep potential well in which it would
oscillate.  These oscillations would lead to a radiative production of
particles.  There have been many variants of the original cold inflation
picture, first introduced in the context of the new inflation model and
shortly afterwards in the chaotic inflation model~\cite{ci}, with many other
models that followed.

The warm inflation picture differs from the cold inflation picture in that
there is no separate reheating phase in the former, and rather radiation
production occurs concurrently with inflationary expansion.  The constraints
by General Relativity for realising an inflationary phase simply require that
the vacuum energy density dominates and so this does not rule out the
possibility that there is still a substantial radiation energy density present
during inflation.  Thus on basic principles, the most general picture of
inflation accommodates a radiation energy density component.  The presence of
radiation during inflation implies the inflationary phase could smoothly end
into a radiation dominated phase without a distinctively separate reheating
phase, by the simple process of the vacuum energy falling faster than the
radiation energy, so that at some point a smooth crossover occurs.  This is
the warm inflation solution to the graceful exit problem.

Dynamically warm inflation is realised if the inflaton were interacting with
other fields during the inflation phase.  In fact in any realistic model of
inflation, the inflaton must be coupled to other fields, since eventually the
inflaton must release its vacuum energy to other fields thereby creating
particles which form the subsequent radiation dominated era in the Universe.
Thus the idea that these couplings to other fields somehow are inactive during
inflation, as pictured in the cold inflation picture, is something that does
require verifying by detailed calculation.  When such calculations are done,
the result is that there are regimes in which particle production during
inflation occurs.  This review will present the calculations which demonstrate
particle production during inflation, thereby leading to a warm inflationary
expansion.

The idea of particle production concurrent with inflationary expansion was
first suggested in the pre-inflation inflation paper by L.Z. Fang in 1980
\cite{Fang:1980wi}.  His paper proposed using a scalar field with the origin
of inflationary expansion due to a claimed anomalous dissipation term that
would be generated based on Landau theory if the field was undergoing a second
order phase transition.  This was dynamically very different from the scalar
field inflation that eventually became successful, and the source of
dissipation was also different from that in warm inflation.  However this
model captured the basic idea of concurrent particle production and inflation.
Then in the mid-80s two papers proposed adding a local $\Upsilon {\dot \phi}$
type dissipation term into the evolution equation of the inflaton field, Moss
\cite{im} and then Yokoyama and Maeda~\cite{Yokoyama:1987an}.  In both cases
the dissipative term generated a source of radiation production during
inflation.  The idea of a dissipative term was re-discovered independently by
Berera and Fang~\cite{Berera:1995wh} almost a decade later.  They went further
by proposing that the consistent dynamics of the inflaton field was a Langevin
equation, in which a fluctuation-dissipation theorem would uniquely specify
the fluctuations of the inflaton field.  That paper by Berera and Fang
provided the foundations for the theory of fluctuations in warm inflation and
the Langevin equation has since been the fundamental equation governing
inflaton dynamics.  {}Following that work, in~\cite{Berera:1995ie} Berera
proposed that a separate reheating phase, as standard in all inflation models
up to then, could be eliminated altogether.  This paper proposed a new picture
of inflation, which it termed warm inflation, in which the process of
inflationary expansion with concurrent radiation production could terminate
simply by the radiation energy over-taking the vacuum energy, thus going from
an inflationary to a radiation dominated era.  This work presented an
alternative solution to the graceful exit problem to the one given by the
standard inflation scenario.  This warm inflation picture was verified
explicitly in \cite{Berera:1996fm}, where the {}Friedmann equations for a
Universe consisting of vacuum and radiation energy were studied and gave many
exact warm inflation solutions to the graceful exit problem. {}Finally
in~\cite{Berera:1999ws} the calculation of fluctuations were done by Berera
for the inflaton evolving by a Langevin equation in a thermal inflationary
Universe.

Alongside the development of the basic scenario, the first principles quantum
field theory dynamics of warm inflation was developed.  This started in
\cite{Berera:1996nv} with a quantum mechanical model which demonstrated the
origin of the fluctuation-dissipation relation in warm inflation.  The key
step in deriving warm inflation from quantum field theory is in realising an
overdamped regime for the evolution of the background inflaton field.  The
initial attempt at this was done by Berera, Gleiser and Ramos in~\cite{BGR}.
In this work, it was proposed that the overdamped evolution should occur under
adiabatic conditions in which the microscopic dynamical processes operated
much faster than all macroscopic evolution, in particular the scalar field
motion and Hubble expansion.  Based on this criteria, a set of consistency
conditions were formulated in \cite{BGR}, which would be required for a
self-consistent solution.  However in~\cite{BGR} no explicit warm inflation
solutions were found.  This point was further highlighted by Yokoyama and
Linde in~\cite{Yokoyama:1998ju}, in which several models were studied from
which the conclusions of~\cite{BGR} were verified.  The problem in these early
works was that dissipation effects were being looked at in a high temperature
regime and it proved too difficult to keep finite temperature effective
potential corrections small, so that the inflaton potential remained
relatively flat, and at the same time obtain a large dissipative coefficient.
One type of model was shown that could realise such requirements~\cite{BGR2}
and this was the first quantum field theory model of warm inflation, although
it was not a very compelling model.  Subsequently Berera and Ramos in
\cite{BR1} suggested a solution for getting around the mutual constraints of
obtaining a large dissipative coefficient and yet small effective potential
corrections.  The main observation was that supersymmetry can cancel local
quantum corrections, such as zero temperature corrections to the effective
potential, whereas temporally non-local quantum effects, such as those that
underly the dissipative effects, will not be cancelled.  This led to \cite{BR1}
proposing a two-stage interaction configuration, in which the inflaton was
coupled to heavy ``catalyst'' fields with masses larger than the temperature of
the Universe and these fields in turn were coupled to light fields.  The
evolution of the inflaton would induce light particle production via the heavy
catalyst fields.  Since these heavy catalyst fields were basically in their
ground state, the quantum corrections associated with them could be cancelled
in supersymmetric models.  The calculation of the low temperature dissipative
coefficients for this two-stage mechanism were first done by Moss and Xiong
\cite{mx}.

There is an earlier review which covered the basics of the warm inflation
scenario \cite{Berera:2006xq}.  In this review full details will be developed
of the quantum field theory dynamics of warm inflation.  This will first start
in Sec.  \ref{wipicture} with a summary of the warm inflation scenario,
including a comparison of it to cold inflation.  In Sec. \ref{sec TFT} a basic
introduction to thermal field theory is given including the real time
formalism for interacting field theories.  In Sec. \ref{effeom} the effective
evolution equation of the inflaton field is derived, in which all fields it
interacts with are integrated out, leading to a Langevin type nonconservative
equation which contains a dissipative term and a noise force term.  The
detailed properties of these dissipation and fluctuation terms is then studied
in Sec.  \ref{fd}.  In addition the physical picture of the dissipation
effects in warm inflation are discussed.  In Sec. \ref{FRWspacetime} the
calculations are extended to curved spacetime.
In Sec.
\ref{particlemod} various particle physics models of warm inflation are
presented.  Finally Sec. \ref{concl} presents some concluding remarks
and future work being done on the subject.
Our conventions are as follows. We use spacetime metrics with $(+---)$
signature, and we use natural units
for the Planck's constant, Boltzmann's constant and the velocity of light
$\hbar=k=c=1$.

\section{The Warm Inflation Picture and its Motivations} 
\label{wipicture}

\subsection{Background equations}

The foundation of cosmology is the cosmological principle which states no
observer occupies a preferred position in the Universe.  This principle
implies the Universe must be homogeneous (looks the same from every point) and
isotropic (looks the same in all directions).  To quantify the cosmological
principle, in regards the geometry, the metric of a space-time which is
spatially homogeneous and isotropic is given by the Friedmann-Robertson-Walker
(FRW) metric,

\begin{equation}
ds^2 =  dt^2 - a^2(t) \left[\frac{dr^2}{1-kr^2}  + r^2 d\theta^2  + r^2 \sin^2
  \theta d \varphi^2 \right],
\label{frw}
\end{equation}
where $(r,\theta,\varphi)$ are spherical-polar coordinates parameterizing the
spatial dimensions, $t$ is cosmic time, and $k=1,-1,0$ describe spaces of
constant positive, negative and zero spatial curvature.  The most important
quantity here is $a(t)$ which is the cosmic scale factor and describes the
expansion of the Universe. The Hubble parameter, defined by

\begin{equation}
H={\dot a\over a}\;,
\end{equation}
quantifies how fast the Universe expands.

The evolution of the scale factor is related to the pressure $p$ and density
$\rho$ by the scale factor acceleration equation,

\begin{eqnarray}
\frac{\ddot a}{a} = - \frac{4 \pi G}{3}(\rho + 3p)\;,
\label{scalee}
\end{eqnarray}
where $G$ is Newton's gravitational constant. Energy conservation is expressed
by
\begin{equation}
{\dot \rho} = -3 H (\rho + p) .
\label{econs}
\end{equation}

Inflation by definition is a phase when the scale factor is growing at an
accelerated rate, ${\ddot a} > 0$, which based on Eq. (\ref{scalee}) requires
$p <- \rho/3$, thus for a substance with negative pressure. The spatial part
of the metric evolves rapidly towards the flat metric $k=0$, and

\begin{equation}
3H^2=8\pi G\rho \;.
\label{fried}
\end{equation}
The most common example of inflationary expansion occurs when the dominant
form of matter has equation of state $p_v = -\rho_v\approx\hbox{constant}$,
which is called vacuum energy.  Such an equation of state plugged into Eq
(\ref{fried}) leads to an exponential scale factor behavior $a(t) \sim
\exp(Ht)$, with $H$ constant.

\subsection{Inflaton dynamics}

\begin{widetext}
\begin{figure}[ht]
  \vspace{1cm} \epsfysize=8.50cm
  {\centerline{\epsfbox{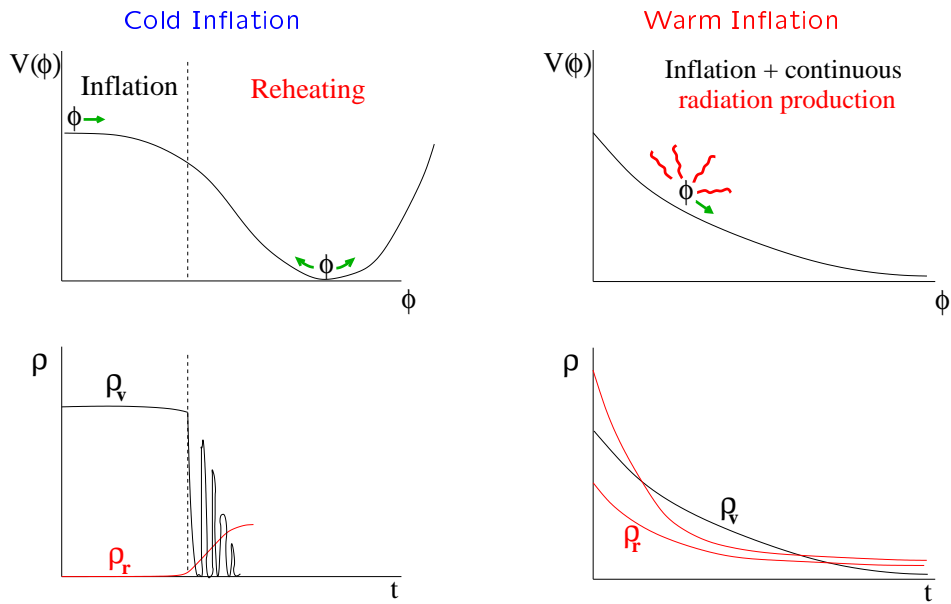}}}
\caption{Comparison of  the cold and  warm inflationary pictures 
\cite{tayfra}.   
Top graphs  show the scalar field evolution and the bottom graphs 
show the vacuum and
radiation energy density evolution.}
\label{ciwi}
\end{figure}
\end{widetext}

The key problem of inflationary cosmology has been trying to realise inflation
from a realistic particle physics motivated model.  The general observation
which has driven this idea is that a scalar field has the necessary equation
of state needed for inflation.  In particular the energy and pressure density
of a scalar field are, respectively, given by

\begin{eqnarray}
\rho  &  =   &  \frac{{\dot  \phi}^2}{2}  +  {\rm   V}(\phi)  +  \frac{(\nabla
  \phi)^2}{2a^2}\;,  \nonumber \\  p  &  = &  \frac{{\dot  \phi}^2}{2} -  {\rm
  V}(\phi) - \frac{(\nabla \phi)^2}{6a^2} ,
\label{rhop}
\end{eqnarray}
with the key point being that the potential energy of this field has precisely
the equation of state conducive to inflation.  Thus, the basic idea has always
been to somehow get the potential energy of a scalar field to dominate the
energy density in the Universe, and thereby drive inflation.  And then, once
enough inflation has occurred, to convert the potential energy into radiation
and enter into a radiation dominated expansion phase.  There are two
underlying dynamical realisations of inflation, into which all models fall,
cold and warm inflation.  Both dynamical pictures are summarized in {}Fig.
\ref{ciwi} and this subsection will review both pictures.

\subsubsection{Cold inflation}
\label{subsect7a}

\begin{figure}[ht]
  \vspace{1cm} \epsfysize=3cm {\centerline{\epsfbox{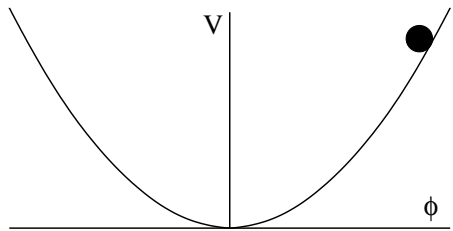}}}
\caption{A  quadratic  inflationary  potential  with  the  inflaton  initially
  starting at some large amplitude.}
\label{figqdp}
\end{figure}

This is the standard picture of inflationary dynamics.  In the cold
inflationary picture, one or more scalar inflaton fields are assumed to
decouple from everything apart from gravity. The energy density is dominated
by the scalar field potential. {}For example, in {}Fig. \ref{figqdp} it is
shown a $m_{\phi}^2 \phi^2$ potential, in which when the inflaton amplitude is
displaced to some $\phi > 0$, inflation can occur.  The evolution of the
scalar field in the FRW universe is described by the General Relativistic
version of the Klein-Gordon equation,

\begin{equation}
{\ddot \phi} + 3H {\dot \phi} - \frac{1}{a^2(t)} \nabla^2 \phi + V'(\phi) = 0.
\label{cieom}
\end{equation}
In this equation the Hubble damping term, $3H {\dot \phi}$, formally acts like
a friction term that damps the inflaton evolution.  However this $3H {\dot
  \phi}$ term does not lead to dissipative energy production, since its origin
is from the coupling of the scalar field with the background FRW metric.  The
inflaton plays the role of driving inflation as well as providing the seeds
for density fluctuations. This Subsection only focuses on the first of these
requirements.

In order for inflation to occur, the inflaton must be potential energy
dominated, which means the potential energy $V(\phi)$ must be larger than the
gradient energy $(\nabla \phi)^2/2$ and the kinetic energy ${\dot \phi}^2/2$,

\begin{equation}
V(\phi) \gg (\nabla \phi)^2/2, {\dot \phi}^2/2 .
\end{equation}
Moreover, in order to obtain enough inflation, these conditions must persist
for some span of time. The usual way to achieve this is for the inflaton to
start out almost homogeneous and at rest in some small patch of space and then
to have the inflaton evolution equation overdamped, with approximate form

\begin{eqnarray}
&&3H\dot\phi+V'(\phi)\approx 0\;,
\label{slowre1}\\ 
&&3H^2\approx 8\pi GV(\phi)\;.
\label{slowre2}
\end{eqnarray}
The consistency of this approximation is governed by conditions on a set of
two slow-roll parameters,

\begin{eqnarray}
\epsilon  &  \equiv  &  \frac{m^2_{P}}{2} \left(\frac{V'}{V}\right)^2  \ll  1,
\nonumber \\ \eta & \equiv & m^2_{P} \frac{V''}{V} \ll 1,
\label{slowci}
\end{eqnarray}
where $m_P^{-2} \equiv 8 \pi G$, so $m_P = 2.4 \times 10^{18}$ GeV. If these
conditions hold in a region of space, then inflation can happen.

{}For the inflaton dynamics described by Eq. (\ref{cieom}), it is instructive
to see how an inflationary scale factor growth occurs. It follows from the
approximate Eqs.  (\ref{slowre1}) and (\ref{slowre2}) that $\dot H=-\epsilon$
and therefore, for $\epsilon \ll 1$ the expansion is roughly constant. {}From
Eq.  (\ref{slowre2}) we see that $a(t)\approx a(0)\exp(Ht)$, where
$H^2\approx8\pi GV/3$, as we stated earlier.
 
The evolution of any radiation contribution to the energy density in the
universe also can be easily studied in this example.  The energy conservation
equation becomes ${\dot \rho_r} = - 4 H \rho_r$, which has the exponentially
decaying solution $\rho_r \sim \rho_r(0) \exp(-4H t)$.  In other words,
whatever the initial radiation energy density in the universe at the onset of
cold inflation, this rapidly decays away, thus supercooling the universe.  As
shown in Fig. \ref{ciwi}, during cold inflation, the vacuum energy density is
large and almost constant, whereas the radiation energy density is negligible,
hence the name cold inflation.

Once a region of adequately large potential energy materializes, the physics
of the subsequent evolution is quite straightforward.  The gravitational
repulsion caused by the negative pressure drives that region into a period of
accelerated expansion.  One expresses the amount of inflation as the ratio of
the scale factor at the end of inflation $a_{EI}$ to that at the beginning
$a_{BI}$, and it is usually stated in terms of the number of e-folds $N_e$,

\begin{equation}
N_e=\ln{a_{EI}\over a_{BI}}.
\end{equation}
In order to inflate the horizon size to a scale covering the observable
universe, it is necessary to have around 60 e-folds of inflation.

Eventually, inflation must end and radiation must be introduced into a very
cold universe so as to put it back into a radiation dominated Hot Big Bang
regime, which is the so called graceful exit problem \cite{oldinf,ni,reheatu}.
In the cold inflation picture, the process that performs this task is called
reheating \cite{reheatu}.  This is usually envisioned as occurring shortly
after the slow-roll approximation has broken down, and is often associated
with oscillations of the inflaton field about the minimum of its potential. An
example of this type of potential 
will be studied in more detail in Subsec.  \ref{workexp}.
If the inflaton is interacting with other matter fields, the oscillations of
the inflaton will lead to particle production so that, as shown in {}Fig.
\ref{ciwi}, the radiation energy density begins to increase.

\subsubsection{Warm inflation}
\label{subsect7b}

The other dynamical realisation of inflation is warm inflation.  This picture
is similar to cold inflation to the extent that the scalar inflaton field must
be potential energy dominated to realise inflation.  The difference is, in
this picture the inflaton is not assumed to be an isolated, non-interacting
field during the inflation period.  So, rather than the Universe supercooling
during inflation, instead the Universe maintains a small amount of radiation
during inflation, enough to noticeably alter inflaton dynamics.  In
particular, the dividing point between warm and cold inflation is roughly at
$\rho_r^{1/4} \approx H$, where $\rho_r$ is the radiation energy density
present during inflation and $H$ is the Hubble parameter. Thus $\rho_r^{1/4} >
H$ is the warm inflation regime and $\rho_r^{1/4} \stackrel{<}{\sim} H$ is the
cold inflation regime. This criterion is independent of thermalisation, but if
such were to occur, one sees this criteria basically amounts to the warm
inflation regime corresponding to when $T > H$.  This condition is easy to
understand since the typical inflaton mass during inflation is $m_{\phi}
\approx H$ and so when $T>H$, thermal fluctuations of the inflaton field will
become important.
Subsequent to the introduction of warm inflation, other scenarios
have been suggested which utilize some of its concepts
such as thermal fluctuations during inflation 
\cite{Alexander:2001dr,Magueijo:2002pg,Ferreira:2007cb}
and graceful exit
via radiation energy density that smoothly becomes 
dominant \cite{Alexander:2001dr};
however the dynamics in these scenarios differs
from that of scalar field stochastic evolution.

The interaction of the inflaton with other fields implies its effective
evolution equation in general will have terms representing dissipation of
energy out of the inflaton system and into other particles.  Berera and Fang
\cite{Berera:1995wh} initially suggested that for a consistent description of
an inflaton field that dissipates energy, the inflaton evolution equation
should be of the Langevin form, in which there is a fluctuation-dissipation
relation which uniquely relates the field fluctuations and energy dissipation.
This has formed the basis of all fluctuation calculations in warm inflation.
The simplest such equation would be one in which the dissipation is temporally
local,

\begin{equation}
{\ddot \phi} + [3H + \Upsilon] {\dot \phi}
- \frac{1}{a^2(t)} \nabla^2 \phi + V'(\phi)= \xi.
\label{wieom}
\end{equation}
In this equation, $\Upsilon {\dot \phi}$ is a dissipative term and $\xi$ is a
fluctuating force.  Both are effective terms arising due to the interaction of
the inflaton with other fields.  In general these two terms will be related
through a fluctuation-dissipation theorem, which would depend on the
statistical state of the system and the microscopic dynamics.  Equations like
(\ref{wieom}) are the main subject of this review, and will be further
described in the next subsection and in Secs. \ref{sec TFT}-\ref{fd}.

In order for warm inflation to occur the potential energy $\rho_v$ must be
larger than both the radiation energy density $\rho_r$ and the inflaton's
kinetic energy.  A major difference from cold inflation is in the evolution of
the energy densities, as can be compared in {}Fig.  \ref{ciwi}.  In warm
inflation the radiation energy does not vanish because vacuum energy is
continuously being dissipated at the rate ${\dot \rho}_v = - \Upsilon {\dot
  \phi}^2$.  The energy conservation equation (\ref{econs}) for this system of
vacuum and radiation becomes

\begin{equation} 
{\dot \rho}_r = -4H \rho_r + \Upsilon {\dot \phi}^2.
\label{radwi}
\end{equation}
In this equation the second term on the right-hand-side acts like a source
term which is feeding in radiation energy, whereas the first term is a sink
term that is depleting it away.  When $H$, $\Upsilon$ and $\phi$ are slowly
varying, which is a good approximation during inflation, there will be some
nonzero steady state point for $\rho_r$.  Thus at large times, compared to the
local Hubble time, the radiation in the universe becomes independent of
initial conditions and depends only on the rate at which the source is
producing radiation.

A series of slow-roll conditions must be satisfied for a prolonged period of
inflation to take place. The slow-roll parameters for warm inflation are

\begin{equation}
\epsilon={m_P^2\over             2}\left({V'\over            V}\right)^2,\quad
\eta=m_P^2\left({V^{\prime\prime}\over                          V}\right),\quad
\beta=m_P^2\left({\Upsilon'V'\over \Upsilon V}\right).
\label{slowrp}
\end{equation}
The slow-roll conditions for warm inflation can be expressed as

\begin{eqnarray}
\epsilon<1+Q,\qquad\eta<1+Q,\qquad \beta<1+Q,
\label{slowwi}
\end{eqnarray}
where the parameter $Q$ is defined by

\begin{equation}
Q \equiv \frac{\Upsilon}{3H}.
\label{uhratio}
\end{equation}
The inflationary solution to the system of equations can be shown to be an
attractor when the slow-roll conditions are satisfied \cite{xiong}.  These
conditions can be weaker than the corresponding slow-roll conditions for cold
inflation if $Q$ is large.  Inflation ends when the vacuum energy ceases to
dominate, which is typically when $\rho_v=\rho_r$ and $\epsilon=1+Q$.  An
example is shown in {}Fig. \ref{ciwi}. Exact solutions for background warm
inflationary cosmologies of radiation and vacuum energy densities were
computed in \cite{Berera:1996fm}.

Additional slow-roll conditions must be imposed if the dissipation coefficient
or the potential depend on the radiation density. For example, in the case of
thermal radiation there are quantum thermal corrections to the inflaton
potential. An additional slow-roll parameter describes this effect,

\begin{equation}
\delta={TV_{,\phi T}\over V_{,\phi}}.
\end{equation}
The slow-roll condition for $\delta$ is stronger than the slow-roll conditions
on the other parameters \cite{xiong},
\begin{equation}
\delta<1.
\end{equation}
This condition is crucial to the realisation of warm inflationary models.
Basically, this condition states that viable warm inflationary models use some
mechanism for suppressing thermal corrections to the inflaton potential.

It is worth stressing the fact that the presence of radiation during the
inflationary epoch is perfectly consistent with the equations of motion. All
that is required is that the vacuum energy density $\rho_v$ be larger than the
radiation energy density $\rho_r$.  In cold inflation, both the radiation
energy density and the friction term are negligible, {\it i.e.},
$\rho_r^{1/4} \ll H$ and $\Upsilon \ll H$.  The main difference between warm
inflation and cold inflation is the reversal of this first condition, i.e.
$\rho_r^{1/4}>H$. There are two regimes that can then be identified, strong
and weak dissipative warm inflation. Strong dissipative warm inflation is the
case $\Upsilon > 3H$, and weak dissipative warm inflation is $\Upsilon \leq 3H$.
The terminology here is almost self-explanatory, in the strong dissipative
regime, the dissipative coefficient $\Upsilon$ controls the damped evolution
of the inflaton field and in the weak dissipative regime, the Hubble damping
is still the dominant term.

Even though the presence of radiation need not hinder inflationary growth, it
can still influence inflaton dynamics.  Consider, for example, inflation
occurring at the Grand Unified Theory scale, which means $V^{1/4} \sim
10^{15}$ GeV.  In this case the Hubble parameter turns out to be $H \sim
V^{1/2}/m_P \sim 10^{10}$ GeV.  {}For cold inflation and weak dissipative warm
inflation, with just the Hubble damping term, the effective inflaton mass
$m_\phi=(V'')^{1/2}$ could be at most $m_\phi \sim 10^{9-10}\, {\rm GeV}<H$.
The key point to appreciate here is that there are five orders of magnitude
difference here between the vacuum energy scale and the scale of the inflaton
mass.  Thus there is a huge difference in scales between the energy scale
$V^{1/4}$ driving inflation and the energy scale $m_\phi$ governing inflaton
dynamics.  This means, for example, in order to excite the inflaton
fluctuations above their ground state only requires a minuscule fraction of
the vacuum energy to be dissipated as radiation, something at a level as low
as $0.001\%$.  This gives an indication that dissipative effects during
inflation have the possibility to play a noticeable role (several models of
warm inflation models exploiting these properties exist
\cite{Berera:1996fm,dr,gmn,Berera:1998hv,mb,lf,BGR2,ml,Billyard:2000bh,tb,Donoghue:2000fk,dj,Dymnikova:2001jy,cjzp,DeOliveira:2002wk,by,hmb,Jeannerot:2005mc,Mimoso:2005bv,ch,bb4,Hall:2007qw,BuenoSanchez:2008nc,Battefeld:2008py,Mohanty:2008ab}). 
Of
course, this energy scale assessment is only suggestive.  This remains a
question that only a proper dynamical calculation can answer.  In particular,
the universe is expanding rapidly during inflation, at a rate characterised by
the Hubble parameter $H$, and one must determine whether the fundamental
dynamics responsible for dissipation can occur at a rate faster than the
Hubble expansion.

Another difference between warm inflation and cold inflation is how the
slow-roll conditions feed back into conditions on the parameters of the
underlying particle model.  This in particular becomes evident in the strong
dissipative regime when $\Upsilon \gg 3H$.  To appreciate this point, note
that in cold inflation the slow-roll conditions require inflaton mass to
satisfy $m_\phi<H$, and this can be a problem in realistic quantum field
theory models of inflation.  The reason most realistic models of inflation
rely on supersymmetry is to help cancel quantum corrections and thus maintain
the desired flatness of the inflaton potential.  However, supersymmetry can be
local as well as global.  Any local, or supergravity theory has associated
with it a K{\" a}hler potential which alters the scalar field potential
\cite{sugra}.  Generic K{\" a}hler potentials lead to inflaton masses bigger
than $H$, thus contradicting the slow roll conditions.  One must then consider
only very special K{\" a}hler potentials that can be consistent with
inflation.  This restricts model building prospects, but even more seriously,
it is very likely that these very special forms may not be produced when the
K{\" a}hler potential is derived, as opposed to put in by hand.  This is often
called the ``$\eta$-problem''
\cite{Copeland:1994vg,Dine,Gaillard:1995az,Kolda:1998kc,Arkani-Hamed:2003mz}.
There are solutions that attempt to
stablize the flat
directions \cite{Gaillard:1995az,Davis:2008sa}
thus allow cold inflation.  Moreover a very different proposal 
that can overcome the ``$\eta$-problem'' in cold inflation is
D-term inflation \cite{Halyo:1996pp,Binetruy:1996xj}.  
However all these solutions require greater
model building details.
In contrast, in warm inflation slow-roll motion only requires from Eq.
(\ref{slowwi}) $\eta<1+Q$, which means that when $\Upsilon > 3H$ the inflaton
mass $H < m_\phi<(H\Upsilon)^{1/2}$, which can be much bigger than in the cold
inflation case.  This relaxation on the constraint in the inflaton mass
permits much greater freedom in building realistic inflaton models, since the
``$\eta$-problem'' is eliminated.

Another model building consequence differing warm inflation to cold inflation
relates to the range of the scalar field $\phi$ in which the inflation occurs.
{}For cold inflation, for the simplest kinds of potentials, which also are the
most commonly used, such as $V = \lambda \phi^4/4$ and $V = m_{\phi}^2
\phi^2/2$, calculations show that the inflaton range has to be above the
Planck scale $\phi > m_P$.  This arises because inflation ends when
$\phi\approx m_P$, and in order to have the desired 60 or so e-folds of
inflation, the inflaton has to start with a value larger than $m_P$.  Although
the potential can still be below the Planck energy density, there are likely to be
difficulties from quantum gravity or supergravity effects, which are discussed in
further detail in Sec.  \ref{particlemod}.  The upshot is that restrictions have to be placed
on model building simply to avoid this problem.  On the other hand in 
warm inflation, when $\Upsilon > 3H$, the added
dissipation means the period of slow roll necessary to obtain the desired
60 or so e-folds can be achieved with the inflaton traversing over a much
smaller range.  For example, with monomial potentials, the inflaton amplitude is 
below the Planck scale $\phi < m_P$ in warm inflation. Potentials which
do not allow cold inflation can sometimes be used for warm inflationary model
building, as will be further discussed in 
Sec. \ref{particlemod}.

\subsection{Fluctuations}

The sources of density fluctuations in warm inflationary models are the
thermal fluctuations in the radiation fields. This is a substantial departure
from cold inflation, where the density fluctuations arise from quantum vacuum
fluctuations.  In this subsection the inflaton fluctuations during warm
inflation are related to a Langevin equation for the inflaton field.  An
intuitive argument is presented and results from the systematic derivation
given.

Inflaton fluctuations are described by the amplitude $\delta \phi ({\bf k},
t)$ for comoving wave number $k$ and cosmic time $t$.  These satisfy a Langevin
equation similar to Eq.  (\ref{wieom}), but now in an expanding universe
\cite{Berera:1995wh,Berera:1999ws},
\begin{equation}
{\delta\ddot \phi} ({\bf k},t) +  (3H + \Upsilon) \delta \dot\phi({\bf k},t) +
({\bf k}^2 a^{-2} + m^2) \delta \phi({\bf k}, t) = \xi({\bf k},t).
\label{langwi}
\end{equation}
In the above equation, the noise correlator is taken to satisfy the
fluctuation-dissipation relation, in which case we have the result

\begin{equation}
\langle \xi({\bf k}, t) \xi({\bf k}', t') \rangle = 2 (3H + \Upsilon) T a^{-3}
(2\pi)^3 \delta^3({\bf k} - {\bf k}') \delta(t-t') .\label{xixi}
\end{equation}
The noise drives scalar field fluctuations with amplitude
$\overline{\delta\phi({\bf k},t)}$, defined by
\begin{equation}
\langle   \delta    \phi({\bf   k},t)   \delta    \phi({\bf   k}',t)   \rangle
=k^{-3}\overline{\delta\phi}^2 (2\pi)^3 \delta^3({\bf k} - {\bf k}') .
\end{equation}

The regime of interest here is where the zero-mode is overdamped, {\it i.e.},
meaning $3H + \Upsilon > m$.  As time progresses, the oscillation frequency
$\omega_k=({\bf k}^2 a^{-2} + m^2)^{1/2}$ decreases until eventually the mode
gets frozen in, similar to what happens in cold inflation.  However, since the
dissipative term in warm inflation can be much larger than that in cold
inflation due to the $\Upsilon$ term, this freeze-out momentum scale can be
much larger than that in cold inflation, which is $\sim H$.  At the freeze-out
time $t_F$, when the physical wavenumber $k_F=k/a(t_F)$, the mode amplitude
$\overline{\delta\phi}$ can be estimated using a purely thermal spectrum,
\begin{equation}
\overline{\delta   \phi}^2(k_F)  \approx   \int_{k<k_F}  \frac{d^3k}{(2\pi)^3}
         {1\over\omega_k}(e^{\beta\omega_k}-1)^{-1}   \stackrel{T  \rightarrow
           \infty}{\approx} \frac{k_F T}{2 \pi^2} .
\label{dphiest}
\end{equation}
Note that, since the solutions to the source-free equation for $\delta\phi$
are heavily damped, the system looses any memory of the initial conditions.

To estimate $k_F$, one must determine when the damping rate of Eq.
(\ref{langwi}) falls below the expansion rate $H$, which occurs at
$k_F^2\approx (3H+\Upsilon)H$.  Thus, in the strong dissipative regime $Q \gg
1$, this implies $k_F \sim \sqrt{H \Upsilon}$.  Substituting for $k_F$ in Eq.
(\ref{dphiest}), one finds the expression for the inflaton amplitude at
freeze-out
\begin{equation}
\overline{\delta \phi}^2 \sim \frac{\sqrt{H \Upsilon} T}{2 \pi^2} .
\label{dphistrong}
\end{equation}
This expression was first derived by Berera in \cite{Berera:1999ws}. 
In the weak
dissipative regime $Q \ll 1$, the freeze-out wavenumber $k_F \sim H$, the
latter being consistent with what occurs in cold inflation. The inflaton
amplitude at freeze-out becomes
\begin{equation}
\overline{\delta \phi}^2 \sim \frac{H T}{2 \pi^2} .
\label{dphiweak}
\end{equation}
This expression was first found by Moss in \cite{im} and then independently
rediscovered by Berera and Fang in \cite{Berera:1995wh}.  
In both cases it was incorrectly
asserted to be the expression for the entire dissipative regime, and in
\cite{Berera:1999ws} the appropriate regime of its validity, the weak
dissipative regime, was clarified.

A much more rigorous description of the fluctuation amplitude can be found in
Hall, Moss and Berera \cite{hmb} and Moss and Xiong \cite{Moss:2007cv}. These
papers solved the Langevin equation (\ref{langwi}) explicitly using Green's
function methods. They also solved the full set of equations for linear
fluctuations, including metric and entropy perturbations, in addition to the
inflaton perturbations. Solving the Langevin equation gives
\begin{equation}
\overline{\delta\phi}^2   \approx   k^{-3}{\sqrt{\pi}\over   2}  \left[(3H   +
  \Upsilon)H\right]^{1/2} T ,\label{dphicorrect}
\end{equation}
at the freezout scale, in agreement with the heuristic description above.
However, a new effect can be seen when the friction coefficient depends on the
temperature of the radiation. In this case the fluctuation amplitude has an
oscillatory dependence on scale, caused by the entropy fluctuations which are
present on sub-horizon scales.

\subsection{Worked example}
\label{workexp}

In this Subsection an example of the quadratic 
potential is presented to show how
inflation models are solved in both the cold and warm inflation dynamics.  The
results will also help illustrate some key features that differentiate the two
dynamics.

\subsubsection{Cold inflation}

We start with the slow-roll equations (\ref{slowre1}) and (\ref{slowre2}) with
potential
\begin{equation}
V(\phi)=\frac12 m_\phi^2\phi^2.
\end{equation}
To solve an inflation model, one must first determine the value $\phi_{EI}$ at
which inflation ends, and then evolve the field backwards to find the value of
the field $N_e$ e-folds before the end of inflation, which we call
$\phi_{N_e}$. The slow-roll parameters Eq. (\ref{slowci}) for this potential
become
\begin{equation}
\epsilon = \eta ={2 m_P^2\over \phi^2},
\end{equation}
so that the slow-roll conditions are satisfied for $|\phi| \gg m_p\sqrt{2}$.
Inflation ends when the slow-roll parameter $\epsilon=1$,
after which the inflaton starts oscillating and the reheating
phase commences.  The number of
e-folds is computed as
\begin{equation}
N_e       =      \int_{t_{N_e}}^{t_{EI}}       H      dt       =-      {1\over
  m_P^2}\int_{\phi_{N_e}}^{\phi_{EI}} \frac{V}{ V'} d\phi.
\end{equation}
Thus, for our quadratic potential we find $\phi_{EI} = m_p\sqrt{2} $ and
\begin{equation}
\phi^2_{N_e} = 2 m_P^2(2N_e +1).
\end{equation}

Next, we choose $N_e$ to correspond to the largest observable scales and fix
the amplitude of density fluctuations to coincide with the observed value. The
amplitude is given by
\begin{equation}
\delta_H = \frac{2}{5} \frac{H \delta \phi}{|{\dot \phi}|} ,
\label{deltah}
\end{equation}
where all quantities are evaluated as the perturbation exits the horizon.
Using the cold inflation expression $\delta \phi =H$ and the slow-roll
approximation, it leads to
\begin{equation}
\delta_H = 0.52 N_e^2\,{m_\phi\over m_P}.
\end{equation}
Setting $\delta_H$ to the observed value $ \approx 2\times10^{-5}$ 
and using $N_e=60$
leads to $m_{\phi} \approx 6.4 \times 10^{-7} m_{P}$.

{}Finally, when computing the spectral index, $n_s$, we find the result

\begin{equation}
n_s -1=2\eta - 6 \epsilon =-{2\over N_e}.
\end{equation}
Note that in this model the inflaton background amplitude $\phi_{60}> m_{P}$
and the expansion rate $H_{60} > m_{\phi}$.  Both these features are common to
such monomial cold inflation models and, as discussed in Sec.
{\ref{particlemod}}, pose model building problems.

\subsubsection{Warm inflation}

Consider strong dissipation $\Upsilon \gg 3 H$ with $\Upsilon$ constant, then
the first slow-roll equation becomes
\begin{equation}
\Upsilon {\dot \phi} + m_{\phi}^2 \phi = 0.
\label{wieom2}
\end{equation}
The number of e-folds of inflation between $\phi_{N_e}$ and $\phi_{EI}$ is now
given by
\begin{equation}
N_e  = \int_{t_{N_e}}^{t_{EI}} H  dt =  -\int_{\phi_{N_e}}^{\phi_{EI}} \frac{H
  \Upsilon}{V'(\phi)} d \phi.
\label{newi}
\end{equation}
Inflation ends when $\epsilon=1+Q$, where $Q$ is defined
in Eq. (\ref{uhratio}). Using the slow-roll
equation  (\ref{newi}) gives $\phi_{N_e} \approx \sqrt{6} (N_e+1) m_{P}
m_{\phi}/\Upsilon$.

Having now determined $\phi_{N_e}$, we can calculate the amplitude for density
perturbations at this point and normalise it to the observational value.
{}For this, the same expression Eq. (\ref{deltah}) for the amplitude is used,
except now $\delta \phi$ is given by Eq.  (\ref{dphicorrect}).  Going through
the calculation, with effective particle number $g_*$, we find
\begin{equation}
\delta_H          \approx          0.18          N_e^{3/8}          g_*^{-1/8}
\left(\frac{\Upsilon}{m_{pl}}\right)^{3/4}.
\end{equation}
Once again setting $\delta_H$ to the observational value 
$\approx 2\times 10^{-5}$ and
setting $N_e \approx 60$ and $g_* \sim 100$, leads to a normalisation
condition, $\Upsilon/m_{P} \approx 1.5 \times 10^{-6}$. (In fact, the number
of e-folds can be as low as 40 in warm inflationary models.)  {}Finally, for
the spectral index, we have the general expression in warm inflation
\cite{hmb}
\begin{equation}
n_s  -  1  =   -\frac{1}{Q}  \left(\frac{9}{4}  \epsilon  -  \frac{3}{2}  \eta
\right)=-{3\over 4 N_e}.
\end{equation}
This is slightly closer to $n_s=1$ spectrum than the value for cold inflation
case.

{}From these results we find that
\begin{equation}
{H_{60}\over m_\phi}\approx{1\over \sqrt{6}}{\phi_{60}\over m_{p}} \approx {60
  m_{\phi}\over \Upsilon}\approx 4\times 10^{7} {m_{\phi}\over m_{P}},
\end{equation}
so that choosing the inflaton mass $m_\phi<10^{-8}m_p$ will mean that the
inflaton mass is bigger than the Hubble parameter, thus eliminating the
$\eta$-problem, and $\phi_{60}$ is below the Planck scale, thus making such
models amenable to particle physics model building.  {}Finally note that this
analysis can be extended to other potentials \cite{tb} and to more general
cases where $\phi$ and $T$ dependence is in both $\Upsilon$ and the inflaton
potential \cite{hmb}.

\section{Thermal Field Theory} 
\label{sec TFT}

We turn now to the main subject of this review, which is to describe how the
dissipative effects which where put into the inflaton dynamics in the earlier
sections have been obtained from microphysical descriptions of quantum field
systems close to thermal equilibrium.  In this section we give a basic
introduction to thermal field theory with the assumption the reader is
familiar with ordinary quantum field theory.  For more extensive reviews of
thermal field theory, please see, for example,
Refs.~\cite{chou,rivers,weert,bellac}.

\subsection{Preliminaries}

In thermal field theory the quantities of interest are ensemble averages of
operator expectation values. The `tried and tested' approach to thermal field
theory involves expressing the observable quantities in terms of propagators
and then applying perturbation theory. This is similar in many respects to
ordinary quantum field theory, except that the emphasis is on the evolution of
operators rather than the scattering matrix. It is best in this context to
regard the system as being always in the `in' state, and as we shall see below
this leads to a richer propagator structure than usual.

We shall be using the Schwinger-Keldysh, or the closed-time path (CTP)
approach \cite{schw} to evaluate ensemble averages. We take a complete sets of
states $\psi_i$ and $\psi_f$ along with a density matrix $\rho$.  The
Schwinger-Keldysh generating function is defined in terms of two source terms,
$J_1$ and $J_2$, by

\begin{equation}
Z[J_1,J_2]=               \sum_{i,f}\langle\psi_{i}|\rho\,\,T^*\exp\left(-i\int
J_2\hat\phi\,\right)|\psi_f      \rangle      \langle\psi_f|\,T\exp\left(i\int
J_1\hat\phi\,\right)|\psi_{i}\rangle\;,\label{zop}
\end{equation}
where $T^*$ denotes time ordering of the operators with the smallest time on
the left.

Ensemble averages of products of $\hat\phi$ can be obtained by differentiation
of the generating functional with respect to $J$, for example, an average like

\begin{equation}
\langle                                      T^*\hat\phi(x_1)\dots\hat\phi(x_r)
T\hat\phi(x_{r+1})\dots\hat\phi(x_n) \rangle \;,
\end{equation}
with both `time ordering' $T$ and `reverse time ordering' $T^*$ is obtained
from $r$ derivatives with respect to $J_2$ and $n-r$ derivatives with respect
to $J_1$. It proves convenient to remove the minus sign in front of $J_2$ by
defining $J^1=J_1$ and $J^2=-J_2$.

{}Four different connected two-point functions can be obtained from the second
derivatives of the generating function,

\begin{equation}
G_{ab}(x,x')=-i{\delta \ln Z\over \delta J^a(x)\delta J^b(x')}\;.
\end{equation}
These can be placed neatly into a $2\times 2$ matrix

\begin{equation}
G_{ab}(x,x')=            \left(\begin{array}{cc}           \langle           T
  \hat\phi(x)\hat\phi(x')\rangle_c&                                     \langle
  \hat\phi(x')\hat\phi(x)\rangle_c\\ \langle \hat\phi(x)\hat\phi(x')\rangle_c&
  \langle T^*\hat\phi(x)\hat\phi(x')\rangle_c
\end{array}\right)\;.
\label{propdef}
\end{equation}
We recognise that $G_{11}$ is the thermal analogue of the {}Feynman propagator
$G_F$. The remaining combinations are the thermal Dyson function $G_{22}$, the
thermal Wightman function $G_{21}$ and its transpose $G_{12}$.  Note that the
two-point functions only depend on the {\it initial} density matrix, and
because of this fact the formalism is sometimes called the `in-in' formalism.

The propagator can be split into real and imaginary parts by introducing the
real anticommutator function $F$ and real spectral function $\rho$, defined by

\begin{eqnarray}
\rho(x,x')&=& i\langle [\hat\phi(x),\hat\phi(x')]\rangle_c\;,
\label{rho}\\
F(x,x')&=&\frac12\langle \{\hat\phi(x),\hat\phi(x')\}\rangle_c\;.
\label{functionF}
\end{eqnarray}
The propagator matrix separates into real and imaginary parts according to

\begin{equation}
G_{ab}(x,x')=              \left(\begin{array}{cc}             F(x,x')-{i\over
    2}\sigma(x,x')&F(x,x')+{i\over        2}\rho(x,x')\\       F(x,x')-{i\over
    2}\rho(x,x')&F(x,x')+{i\over 2}\sigma(x,x')
\end{array}\right)\;,
\label{G11Fr}
\end{equation}
where $\sigma(x,x')=\rho(x,x'){\rm sgn}(t-t')$ is the real and time-symmetric
Wheeler-Feynman propagator.

\subsection{Thermal equilibrium}

Systems in thermal equilibrium are invariant under time translation, space
translation and additionally the propagators satisfy periodicity relations in
imaginary time. The propagators depend on $x$ and $x'$ only in the combination
$x-x'$, and it usually proves convenient to use the {}Fourier transform over
space and time, replacing $x-x'$ with $({\bf p},\omega)$.

The imaginary-time periodicity relations imply that the anticommutator
function $F$ and the spectral function $\rho$, in space-time momentum
representation, are related by \cite{bellac}

\begin{equation}
F({\bf p}, \omega) = -\frac{i}{2} \left[ 1 + 2 n(\omega) \right] \rho({\bf p},
\omega)\;,
\label{frhofdt}
\end{equation}
where $n(\omega)$ is the thermal distribution function for inverse temperature
$\beta$,

\begin{equation}
n(\omega)={1\over e^{\beta\omega}- 1}\;.
\end{equation}
This remarkable relation between $F$ and $\rho$, means that the full thermal
propagator depends only on the spectral functions.

The spectral functions can be obtained perturbatively by solving the
Schwinger-Dyson equation,

\begin{equation}
(\omega^2-{\bf k}^2-m^2)G_{ab}-\Sigma_a{}^cG_{cb}=ic_{ab}\;,
\label{sde}
\end{equation}
where $\Sigma_{ab}$ is the self-energy matrix. The tensor $c_{ab}$ is a
diagonal matrix with entries $\pm 1$. It is used as a metric to raise indices
$a$ and $b$, and keeps track of the minus signs introduced by the reverse time
ordering.

As with the propagator, the self-energy matrix can also be expressed in terms
of two functions $\Sigma_F$ and $\Sigma_\rho$,

\begin{eqnarray}
i\Sigma_\rho&=&     i\left(\Sigma_{21}-\Sigma_{12}\right)\;,\\    i\Sigma_F&=&
\frac12\left(\Sigma_{21}+\Sigma_{12}\right)\;.
\label{Sigmadef}
\end{eqnarray}
In thermal equilibrium, the {}Fourier transforms of $\Sigma_\rho$ and
$\Sigma_F$ are related by a local relation just like Eq. (\ref{frhofdt}),

\begin{equation}
\Sigma_F({\bf  p}, \omega)  =-  \frac{i}{2}  \left[ 1  +  2 n(\omega)  \right]
\Sigma_\rho({\bf p}, \omega)\;.
\label{Sfrhofdt}
\end{equation}
The physical interpretation of $\Sigma_\rho$ is that it is related to the
decay processes and can be associated with a relaxation time defined by
\begin{equation}
\tau({\bf     p},     \omega)={4\omega\over     i     \Sigma_\rho({\bf     p},
  \omega)}.\label{relax}
\end{equation}
Equation (\ref{Sfrhofdt}) will turn out later to be related to a
fluctuation-dissipation theorem.

The real time formalism extends to Dirac spinors in a routine manner.  A
description of the thermal propagator for Dirac fields can be found in
\cite{bellac}. The propagator in similar conventions to the ones used here can
be found in \cite{mx}.

\subsection{Real time formalism for interacting field theories}

We now explain how to relate the general Green's functions to the free field
case using perturbation theory, {\it i.e.}, Feynman diagram expansion, in the
context of the Schwinger-Keldysh formalism.  The causality properties of the
{}Feynman diagrams are also discussed.

In path integral form, the generating function (\ref{zop}) becomes

\begin{equation}
Z[J_1,J_2]=\int d\mu[\phi_1]d\mu[\phi_2]\,\rho[\phi_1(t_i),\phi_2(t_i)]\, \exp
\left\{iS[\phi_1]+i\int J_1\phi_1-iS[\phi_2]-i\int J_2\phi_2 \right\}.
\end{equation}
where $\rho$ is the density matrix at the initial time $t_i$ and the paths
cross asymptotically as $t\to\infty$. This path integral is equivalent to
using a single field on a closed time path (CTP), taking the time integration
along a contour in the complex time plane going from $t=t_i$ to $+ \infty$
(forward branch) and then back to $t=t_i$ (backwards branch). We still refer
to the formalism as the CTP formalism even though we find it convenient to
keep the fields and branches distinct.

The CTP {}Feynman diagram expansion for ensemble averages mirrors the ordinary
{}Feynman diagram expansion for $n$-point functions very closely. An important
difference is that vertices carry an extra integer label taking the value $1$
or $2$ which determines the component of the propagator matrix between them.
The vertices represent the difference of interaction Lagrangians, ${\cal
  L}_I(\phi_1)-{\cal L}_I(\phi_2)$, and therefore a minus sign is included for
each vertex labeled $2$. Troublesome minus signs can be removed from the
source terms by using a metric $c^{ab}={\rm diag}(1,-1)$.

One useful feature of the {}Feynman diagram expansion is the {\sl maximum time
  rule}.  If the diagrams are drawn in configuration space, then a diagram
gives a zero contribution if the time on any internal vertex is larger than
the largest external time.  This rule follows from the symmetries of the
Green's function. It implies that the effects of source terms always satisfy
the rules of causality, event though the Green's function is acausal.

The CTP effective action also follows by direct analogy with the usual
effective action,
\begin{equation}
e^{i\Gamma[\phi_1,\phi_2]}=\int_{1PI}
d\mu[\phi_1']d\mu[\phi_2']\,\rho[\phi_1'(t_i),\phi_2'(t_i)]\,  \exp
\left\{iS[\phi_1'+\phi_1]-iS[\phi_2'+\phi_2]\right\}.\label{effectact}
\end{equation}
where only the 1-particle irreducible (1PI) diagrams contribute. It satisfies
the effective field equation
\begin{equation}
\left.{\delta \Gamma\over \delta\phi_a}\right|_{\phi_1=\phi_2}=0.
\end{equation}
The new features introduced in the CTP approach are the doubling of fields and
the condition $\phi_1=\phi_2$.

The path integral may be constructed from the original vertices and Green's
functions, or by shifting the Lagrangian about a background field and then
using shifted vertices and corrected Green's functions.  When the propagators
and the vertices of the {}Feynman diagram expansion depend on the background
fields, they must be kept distinct.  However, after taking the variation of
the effective action to obtain the effective field equations, the values of
$\phi_1$ and $\phi_2$ are set to the same value. Note that the maximum time
rule quoted above only applies when $\phi_1=\phi_2$.

An important case of the background field approach is when the system is close
to thermal equilibrium and the background fields vary slowly compared to the
relaxation time of the system. In this case, it can be appropriate to take the
propagator of the background field expansion to be in thermal equilibrium, and
relate all of the non-equilibrium effects to the background field dependent
interaction terms.

\section{The Effective equations of motion} 

\label{effeom}

Now we introduce effective equations of motion, where there is a background
field, which makes the role of a system, and quantum fields, considered as the
environment to which the system is coupled to. By integrating out the quantum
fields it is possible to arrive at a background field equation, which is of a
Langevin-like type typical of a system in interaction with an environment.  We
will keep our formalism as simple and general as possible, allowing then to
extend it to specific model examples later on.

\subsection{Historical background and motivation}

The basic motivation for the study of the nonequilibrium dynamics of a
background field in the context of a separation between a system and an
environment to which it is coupled to comes from many different physical
systems of interest.  In the case of inflation, it is the inflaton field whose
slow-roll dynamics generates the necessary conditions for a successful
inflationary scenario.  Almost all models of inflation involve evolution with
loss of energy of the inflaton field to other fields (or degrees of freedom)
to which it is coupled to.  This process of energy transfer from the inflaton
(regarded as the system) to the other fields is a fundamental requirement of
equipartition, where some portion of the system's energy has to flow
irreversibly to the environment. This process is not exclusive to inflation,
and occurs in any phase transition where some order parameter characterising
the global thermodynamic properties of the system relaxes to an equilibrium
point.

The study of the nonequilibrium dynamics of fields have been approached using
different techniques of quantum field theory and quantum statistical
mechanics, including variational techniques \cite{eboli}, the use of
resummation techniques, for example using the two-particle irreducible (2PI)
procedure \cite{2pi}, and the use of kinetic equation methods \cite{kinetic}.
These different approaches have in common the possibility of keeping all the
dynamics unitary, but at the expense of keeping track of every field, and not
just the one whose dynamics we are most interested in. Even though this study
might be done in a relatively complete fashion for some simple models, it gets
quickly cumbersome as the number of fields (and field modes) increases.

In the Langevin-like approach we focus on the dynamics of the relevant field
describing the system and not on the the remaining field modes.  These instead
act on the system through dissipative and stochastic noise terms, whose
effects then become manifest.  This is a much more economical way of studying
the nonequilibrium dynamics, since we concentrate only on a given (more
relevant) field mode.  This approach has a long history which can be traced
back to Einstein's famous explanation of Brownian motion \cite{einstein}, and
includes classics such as the work of Caldeira and Leggett \cite{caldeira}. In
their work an harmonic oscillator, regarded as the system, is coupled linearly
to a set of other oscillators, regarded as the thermal bath or environment. By
integrating out the thermal bath, the resulting dynamics of the system becomes
explicitly of Langevin form, where dissipation and stochastic noise emerges.
Extension of these studies to the context of nonlinear couplings between
system and thermal bath have been implemented by Hu, Paz and Zhang
\cite{hu-nl}.

Though in quantum mechanics the system, which is out of equilibrium, and the
thermal bath, which drives the system towards equilibrium, may be well
separated, in the context of nonlinear field theories this distinction may be
considered somewhat blurred.  Even so, for self-interacting field theories,
there are situations where short wavelength modes can serve as the thermal
bath driving the longer wavelength modes, which have slower dynamics, into
equilibrium. In this sense, the field can be its own thermal bath. Of course,
other fields coupled to a background scalar field (the system) may also serve
as the thermal bath.  One of the first implementations of this interpretation
in the context of quantum field theory, and motivated by the reheating problem
in inflation, was the work by Hosoya and Sakagami \cite{hosoya1}, who obtained
an approximate dissipation term in the equation of motion for a scalar field.
They did this by examining small deviations from equilibrium in the Boltzmann
equation for the number density operator and then supplemented this by a
computation of transport coefficients using Zubarev's method for
nonequilibrium statistical operators \cite{hosoya1}.  The derivation of
dissipation terms in the context of the $\lambda \phi^4$ model was also
performed using operator methods by Morikawa and Sasaki in \cite{morikawa1}.
Later, in the context of the CTP formalism, Morikawa \cite{morikawa2} obtained
an effective Langevin-like equation for a scalar field interacting with a
fermionic bath, including explicit fluctuation and dissipation terms.

Other work that analysed the emergence of dissipation and fluctuation in the
effective field dynamics was the work done by Hu and collaborators \cite{hu1},
who analysed a scalar field quantum bath quadratically coupled to a background
scalar field system, while Lee and Boyanovsky considered the case of a scalar
field thermal bath linearly coupled to a background scalar field system
\cite{boya1}, with more realistic couplings considered later on \cite{boya2}.
In the work of Gleiser and Ramos \cite{GR}, a systematic study in the context
of the loop expansion at high temperature was performed for both the $\lambda
\phi^4$ model and also for quadratic coupling to another scalar field.  One
important aspect of the dynamics that was demonstrated by these first
references was that the noise terms emerging in the effective dynamics was in
general colored, {\it i.e.} non Markovian, and multiplicative, {\it i.e.}
field dependent (unless the coupling between system and bath was linear).
Thus, these studies have shown that the effective equation that describes the
approach to equilibrium of the slower moving modes can be quite different from
the typical phenomenological Langevin equation with its white and additive
noise terms.

All these early studies were performed in the context of Minkowski spacetime.
{}One of the first to consider dissipative dynamics in the context of a curved
spacetime was Ringwald \cite{ringwald}. Later on, the problem of dissipation
and damping in a de Sitter spacetime was considered by the authors of Ref.
\cite{boya3}, while more recently the nonequilibrium dynamics of the inflaton
field in the Friedmann-Robertson-Walker spacetime was considered by Berera and
Ramos in \cite{BRfrw}, where an extensive analysis of the dissipation kernels,
entropy and particle production were performed.  In this section and in the
next one we will be most concerned with the dynamics in Minkowski spacetime,
while in Sec.  \ref{FRWspacetime} we will discuss the changes necessary to
implement in order to describe our results in the context of the
Friedmann-Robertson-Walker curved spacetime.

\subsection{Introducing the Keldysh Representation}

We start by noting, from the definitions given in Sec.  \ref{sec TFT} for the
four two-point functions defined in the CTP formalism, Eqs. (\ref{G11Fr}),
that they are not all independent.  This indicates that we can define a linear
transformation of the fields to make some components of the propagator matrix
vanish.  In terms of the field $\phi$ defined in the forward branch of the CTP
contour $\phi_1$ and in the backwards branch $\phi_2$, this linear
transformation (also called Keldysh rotation) leads to two new fields $\phi_c$
and $\phi_\Delta$ defined by

\begin{eqnarray}
&&\phi_{c} = \frac{1}{2}(\phi_1 + \phi_2)\;,
\label{phic} \\
&& \phi_{\Delta} = \phi_1 - \phi_2 \;.
\label{phiDelta}
\end{eqnarray}
Now, it is a matter of simple algebra to show that the propagator matrix
transforms to ($a',b'=c,\Delta$)

\begin{equation}
G_{a'b'}= \pmatrix{F(x,x') &G_{R}(x,x') \cr G_{A}(x,x') & 0}\;,
\label{GRAK}
\end{equation}
where

\begin{eqnarray}
&&G_{R}(x,x')=- i \rho(x,x') \theta(t-t')\;,
\label{GR}\\
&&G_{A}(x,x')=i \rho(x,x') \theta(t'-t)\;,
\label{GA}
\end{eqnarray}
$G_R$ and $G_A$ can be identified with the retarded and advanced two-point
functions, while the function $F(x,x')$, defined in Eq. (\ref{functionF}), is
sometimes also called the Keldysh two-point function.

The self-energy matrix in the Keldysh representation also has a simple form,
\begin{equation}
\Sigma^{a'b'}=\pmatrix{
  0&\Sigma_A(x,x')\cr\Sigma_R(x,x')&-i\Sigma_F(x,x')\cr},\label{sem}
\end{equation}
where
\begin{eqnarray}
\Sigma_R&=&\Sigma_\rho(x,x')\theta(t-t'),\label{srsr}\\ 
\Sigma_A&=&-\Sigma_\rho(x,x')\theta(t'-t).
\end{eqnarray}

Let us see some of the advantages of working with the Keldysh representation for
the fields. As an example consider the classical action for a $\lambda \phi^4$
theory in the CTP formalism,

\begin{eqnarray}
S[\phi_1,\phi_2] &=& \int d^4 x  \left[ \frac{1}{2} \phi_1 \left( - \partial^2
  - m^2  \right)  \phi_1 -  \frac{\lambda}{4  !}   \phi_1^4 \right]  \nonumber
\\ &-&\int d^4  x \left[ \frac{1}{2} \phi_2 \left( -  \partial^2 - m^2 \right)
  \phi_2 - \frac{\lambda}{4 !} \phi_2^4 \right]\;,
\label{S12}
\end{eqnarray}
which in terms of $\phi_c,\phi_\Delta$ becomes

\begin{eqnarray}
S[\phi_c,\phi_\Delta] &=& \int d^4 x  \left[ \phi_\Delta \left( - \partial^2 -
  m^2 \right)  \phi_c -  \frac{\lambda}{4 !} \left(  4 \phi_\Delta  \phi_c^3 +
  \phi_\Delta^3 \phi_c \right) \right]\;.
\label{ScD}
\end{eqnarray}
The first thing to note from the action $S[\phi_c,\phi_\Delta]$ is that it
vanishes for a field configuration that it is the same on the forward and
backwards branches of the closed-time path, i.e.  when $\phi_\Delta=0$.
Although this seems obvious for the classical action, it is important to
realise that the same structure remains true for the effective action, with no
terms independent of $\phi_\Delta$ appearing at any perturbative order.  This
restricts the form of vertex and self-energy corrections. For example, the
$c-c$ term in the self energy has to vanish because of the form of the
self-energy matrix, Eq.  (\ref{sem}).  Another consequence of using the
representation of the action in terms of the $\phi_c$ and $\phi_\Delta$ fields
is that the classical field equation can be obtained from

\begin{equation}
\frac{\delta        S[\phi_c,\phi_\Delta]}{\delta        \phi_{\Delta}       }
\Bigr|_{\phi_\Delta=0} = 0\;,
\label{eom}
\end{equation}
which for Eq.  (\ref{ScD}) can be seen to immediately reproduce the usual
classical equation of motion for a $\lambda \phi^4/4!$ theory,
 
\begin{equation}
(\partial^2 +m^2)\phi_c + \frac{\lambda}{3!} \phi_{c}^3 =0\;.
\label{ceom}
\end{equation}

The down-side of working with the Keldysh representation is that the
simplicity of the propagators is offset by the complexity of the vertices in
the Feynman diagram expansion. The best approach often is to use the original
matrix propagator of the Feynman diagram expansion and then to use the Keldysh
representation for the end result.

\subsection{The Effective Action and Equation of Motion: 
  A System-Environment (Langevin) Interpretation}
\label{seclang}

Due to the special properties of the Keldysh representation, the quantum
corrections to the classical action, {\it i.e.} the effective action Eq.
(\ref{effectact}), can be represented generically in the form:

\begin{equation}
\Gamma[\phi_c,\phi_\Delta] = -\int\,d^4x\,  {\cal F}(x)
\phi_\Delta(x)+\frac12\int\,d^4x\,               d^4x'\,\phi_\Delta(x)\,
i\Sigma_F(x,x')\,\phi_\Delta(x')+O(\phi_\Delta^3) \;,
\label{GammacD}
\end{equation}
where we have kept terms up to second order in the field $\phi_\Delta$.  As we
are going to see next, these terms have an important interpretation in the
definition of the effective dynamics for a background field.

The first term in Eq. (\ref{GammacD}) can be recognized as the term which
leads to the effective field equation ${\cal F}=0$, in the absence of the
second term
in that equation, where

\begin{equation}
{\cal F}(x)=
-\left.{\delta\Gamma[\phi_c,\phi_\Delta]\over\delta\phi_\Delta}\right|_{\phi_\Delta=0}
\;.
\end{equation}
Perturbatively, this is given by the classical field equations plus
corrections from the 1-particle irreducible Feynman diagrams. For the moment,
we shall drop the vertex corrections but keep the full the self-energy. (A
better approximation scheme is adopted in the next section.)  The effective
field equation becomes

\begin{equation}
{\cal   F}=\left[\partial^2+m^2+{\lambda\over   3!}\phi_c^2(x)\right]\phi_c(x)
+\int d^4x'\,\Sigma_R(x,x')\,\phi_c(x') = 0 \;,
\end{equation}
where the self-energy term $\Sigma_R(x,x')=\Sigma_\rho(x,x')\theta(t-t')$. An
important property of the effective equations of motion in the Keldysh
formalism is that the causality is always explicit, a fact reflected here in
the use of the retarded combination of self-energy.

The second term in Eq.  (\ref{GammacD}) is a purely imaginary term in the
effective action that depends only on the self-energy. This term contains
information which is needed to describe fluctuations about the solutions to
the effective field equations.  A useful trick has been developed which
replaces the quantum fluctuations by statistical fluctuations in an ensemble
of random fields.  This is done by performing a Hubbard--Stratonovich
transformation in the functional partition function, introducing a random
field $\xi(x)$ to decouple the quadratic term in $\phi_\Delta$ in Eq.
(\ref{GammacD}).

Consider a functional integral with the classical action replaced by the
effective action (\ref{GammacD}). The tree diagram contributions to this
functional integral generates the full $n$-point functions, just as it does in
the non-CPT formalism.  The quadratic term in $\phi_\Delta$ in the functional
integrand can be written as

\begin{eqnarray}
&&  \exp   \left\{  -{1\over   2}  \int  d^4   x  d^4  x'\,   \phi_\Delta  (x)
  \Sigma_F[\phi_c]   (x,x')   \phi_\Delta(x')   \right\}   \nonumber   \\   &&
  =|\det\Sigma_F|^{1/2} \int D \xi \; \exp \left\{ -\frac{1}{2} \int d^4 x d^4
  x'\,  \xi   (x)  \Sigma_F^{-1}(x,x')   \xi  (x')  +   i\int  d^4   x  \xi(x)
  \phi_\Delta(x) \right\}\;,
\label{noise}
\end{eqnarray}
The functional integrand now has a real quadratic term in the field $\xi(x)$,
and a linear term in $\phi_\Delta$. Taking the tree diagram contributions to
the functional integral with this new effective action gives a stochastic
equation of motion,

\begin{equation}
\left[\partial^2+m^2+{\lambda\over     3!}\phi_c^2(x)\right]\phi_c(x)    +\int
d^4x'\,\Sigma_R(x,x')\,\phi_c(x')= \xi(x) \; .
\label{langevin}
\end{equation}
Equation (\ref{langevin}) can be seen as a Langevin-like equation of motion.
{}From Eq. (\ref{noise}), $\xi(x)$ can be interpreted as a Gaussian stochastic
noise with the general properties of having zero mean, $\langle \xi(x) \rangle
=0$, and two-point statistical correlation function

\begin{equation}
\langle \xi(x) \xi(x') \rangle = \Sigma_F (x,x')\;.
\label{noisetwopoint}
\end{equation}
Statistical averages are defined as functional integrals over the $\xi(x)$
field.
An important property of Eq.  (\ref{langevin}), related to the Langevin-like
form, is the existence of a dissipative-like term. {}To demonstrate this, note
that we can define a dissipation kernel ${\cal D}(x,x')$ as
\cite{yoko,localnoise}

\begin{equation}
\Sigma_\rho(x,x')= - \frac{\partial}{\partial t'} {\cal D} (x,x')\;,
\label{disskernel}
\end{equation}
and Eq. (\ref{langevin}) then becomes,

\begin{eqnarray}
 \left[\partial^2 + m^2 +  \frac{\lambda}{3 !} \phi_c^2(x) \right] \phi_{c}(x)
 + \int d^4 x'  {\cal D} (x,x') \dot{\phi}_c(x')= \xi(x)
\label{langevinphi} \;.
\end{eqnarray}
Under a space-time {}Fourier transform, using the definition
(\ref{disskernel}), we find that the noise kernel given by Eq.
(\ref{noisetwopoint}) and the dissipation kernel ${\cal D} (x,x')$ in Eq.
(\ref{langevinphi}) are related by the relation (\ref{Sfrhofdt}),

\begin{equation}
\Sigma_F({\bf    p},\omega)=   2\omega\left[n(\omega)+{1\over   2}\right]{\cal
  D}({\bf p},\omega)\;.
\label{chifdr}
\end{equation}
 
Note that, in the Rayleigh-Jeans regime $\omega \ll T$, $2 \omega \left[
  n(\omega) + 1/2 \right] \to 2 T$, so that Eq.  (\ref{chifdr}) and Eq.
(\ref{noisetwopoint}) reproduce the classical relation between the fluctuation
two-point function and the dissipation,

\begin{equation}
\langle\xi({\bf  p},t)\xi({\bf  p},t')\rangle= 2T\int{d\omega\over  2\pi}{\cal
  D}({\bf p},\omega)e^{i\omega(t-t')}\;,
\label{clnt}
\end{equation}
which forms the basis of Eq. (\ref{xixi}) used in Sec. II.  In Ref.  \cite{GR}
was demonstrated that, at high temperatures (typically $T\gg m_\chi,m_\phi$),
the noise and dissipation kernels tend to approach local forms and then Eq.
(\ref{langevinphi}) becomes Markovian, with a white multiplicative noise term.
Improved analysis for the localization of the noise, and consequently the
dissipation term, was recently done in Ref. \cite{localnoise}.

\section{Markovian Approximation for the Noise and Dissipation Kernels}
\label{fd}

The possibility of approximating nonlinear and nonlocal equations of motion
like Eq. (\ref{langevinphi}) in a local form offers many advantages.
Typically, solving nonlinear stochastic equations of the form of Eq.
(\ref{langevinphi}) is very hard both analytically and numerically.  Any
method attempting to solve these kind of equations requires keeping the memory
of the past history of the scalar field configuration at each stage of the
evolution.  These equations also typically involve highly oscillating nonlocal
kernels that can lead to errors which quickly build up and that are too hard
to control, thus preventing any simple numerical solution.  There is an
immense saving of effort as well as a much better understanding of the physics
from a local equation as opposed to a nonlocal one, since the former can
generally be analysed with a more transparent numerical treatment than the
latter.  {}For these reasons, attempts have been made to express the equations
in a local or Markovian approximate form.  In particular, it has been
suggested in \cite{BGR} that at high temperatures and with a large set of
heat-bath fields, the existence of many decay channels could lead to an
approximate local Langevin equation of motion for $\phi$. The results in
\cite{BGR} motivated one of the first microscopically motivated models for
warm inflation \cite{BGR2}. The large set of heat-bath fields proposed in
\cite{BGR2} would constitute of a tower of massive modes, in a string
motivated model, through which the inflaton field could interact. In the model
proposed in \cite{BGR2} enough radiation would be produced, leading to an
overdamped motion for the inflaton and making possible to sustain inflation
long enough.

Later attempts to treat the nonlocal kernels, and not relying on a local
approximation, but still at high temperature with a large thermal bath, where
proposed in Refs. \cite{BR1}. There the analysis were based on the possibility
of constructing models where the kernels exhibit a strong exponential damping
in time, making possible a numerical approach to analyse the nonlocal equation
of motion for the scalar field background. The results in Refs. \cite{BR1}
have also shown that a local approximation for the kernels, and then for the
equation of motion, is in very good agreement with the full numerical solution
of the nonlocal equation.  However, as shown in Ref. \cite{mx}, there are
cases where such a strong damping behavior for the kernels are not possible
and, thus, we must resort to other alternative analysis to determine how good
a local approximation is for the nonlocal equation.

Local approximations for equations of motion of the form of Eq.
(\ref{langevinphi}) have been criticized by a number of authors. Lawrie
\cite{lawrie}, for example, has studied the non-equilibrium dynamics using
various approximations to the propagators. He has argued, from a formal point
of view, that the local approximation would violate some specific sum rules in
the kinetic equation approach. In the cases where the local approximation was
tested against the numerical solutions coming from the kinetic equations
derived in \cite{lawrie}, it was found that the local approximation tended to
over-estimate the real dissipation.  A similar conclusion was reached by Aarts
and Tranberg \cite{Aarts:2007ye}, who used a numerical code to evolve the
propagators in a large `$N$' approximation for various models which resemble
those used in warm inflation.

The main drawback with these numerical approaches so far is that they have, of
necessity, to be based on models in which warm inflation is not expected to
occur even in the close-to equilibrium approximation. This is because they
have no mechanism to suppress thermal corrections to the inflaton potential
and, as we argued in Sect \ref{subsect7b}, warm inflation cannot take place.
The warm inflation models discussed in Sec. \ref{particlemod} 
all use some form of
supersymmetry, and at the present time no fully non-equilibrium calculation has
been possible due to the complexity of the field content.

Very recently, in Ref. \cite{localnoise}, specific conditions were derived for
the validity of adopting a local approximation for the dissipation and
fluctuation kernels in a specific model favored by warm inflation and that can
be physically realised in the context of supersymmetric models
\cite{mx,BRplb1,BRplb2}.  In this model, first identified in \cite{BR1}, the
background scalar field $\phi_c$ is coupled to heavy intermediate quantum
fields which in turn are coupled to the light quantum fields.  The dynamics of
dissipation and radiation production in this model is realised by a two-stage
mechanism: the background scalar field indirectly induces particle production
in the light fields through the intermediate heavy fields which in a sense
help to catalyze the effect.  A throughout study of model realisations of such
mechanism for dissipation, in systems near thermal equilibrium has been given
in Ref. \cite{mx}, while application of these effects to inflation have been
shown to have significant importance \cite{bb4}.

One important result drawn from the analysis of Ref. \cite{localnoise} was the
demonstration of the nonexistence of a local approximation for the dissipation
and fluctuation kernels at zero temperature. This result was shown to be a
direct consequence of the application of the Markovian approximation to the
generalised fluctuation-dissipation relation like Eq. (\ref{chifdr}). This in
particular implies that there should be no local first order time derivatives
in the equation of motion like Eq. (\ref{langevinphi}) at $T=0$. This result
was also explicitly shown to be the case in Ref. \cite{mx} (though it was also
shown that higher order local but non analytic derivative terms were
possible).

\subsection{Dissipative Effects: Local Approximation}

As discussed above, the dissipative term takes a local form when the
background field is slowly varying and the system remains close to thermal
equilibrium.  The dissipation in this case is related to the transport
coefficient which we have been calling the friction coefficient $\Upsilon$. We
now give a general formula for the friction coefficient.

Consider the effective field equation for the $\phi$ field in the CTP
formalism, which was generated by
\begin{equation}
{\cal F}(x)=-\left.{\delta\Gamma[\phi_c,\phi_\Delta]\over\delta\phi_\Delta(x)}
\right|_{\phi_\Delta=0}\;.
\end{equation}
In the following, let us explicitly consider that $\phi_c$ is spatially
homogeneous and that it varies slowly about its value $\phi(t)$ at a fixed
time $t$. Set $\delta\phi_c=\phi_c-\phi(t)$ and expand ${\cal F}$ by

\begin{equation}
{\cal F}(x)=\sum_{n=0}^\infty {\cal F}_n(x)\;,
\end{equation}
where

\begin{equation} 
{\cal        F}_n(x)=-{1\over         n!}\int        d^4x_1\dots        d^4x_n
\left.{\delta^{n+1}\Gamma\over\delta\phi_\Delta(x)
  \delta\phi_{c}(x_1)\dots\delta\phi_{c}(x_n)}\right|_{\phi_a=\phi(t)}
\delta\phi_{c}(x_1)\dots\delta\phi_{c}(x_n) \;.
\label{fn}
\end{equation}
The first term ${\cal F}_0$ represents the part of the field equations which
contains no derivative terms, and can be expressed as the derivative of an
effective potential $V(\phi)$,
\begin{equation}
-\left.{\delta\Gamma[\phi_c,\phi_\Delta]\over\delta\phi_\Delta(x)}
\right|_{\phi_a=\phi_t}={\partial V\over \partial \phi}.
\end{equation}

The next term ${\cal F}_1$ depends on the equilibrium self-energy of the
inflaton with constant values of the background field,

\begin{equation}
-\left.{\delta^2\Gamma\over\delta\phi_\Delta(x)\delta\phi_c(x')}
\right|_{\phi_\Delta=0} =(\partial^2+m^2)\delta(x-x')+\Sigma_R(x,x')\;,
\end{equation}
where $\Sigma_R=\Sigma_\rho\theta(t-t')$, as before. The total field equation
up to first order in $\delta\phi$ becomes

\begin{equation}
\ddot\phi+  \int d^4x_1\,  \Sigma_R(x-x_1)  \delta\phi(t_1)+ {\partial  V\over
  \partial\phi}=0.
\end{equation}
We re-iterate that, when the self-energy is calculated, we can take $\phi$
to be constant. Since we expand about thermal equilibrium, we have used the
fact that $\Sigma_R(x,x_1)\equiv\Sigma_R(x-x_1)$.

The non-local dissipative term can be localized when there is a separation of
timescales in the system.  Suppose, for example, that the self-energy
introduces a response timescale $\tau$. If $\phi$ is slowly varying on the
response timescale $\tau$, then we can use a simple Taylor expansion and write
\begin{equation}
\phi(t_1)=\phi(t)+(t_1-t)\dot\phi(t)+\dots
\end{equation}
The $\phi$ equation of motion including the linear dissipative terms is then
\begin{equation}
\ddot\phi+\Upsilon\dot\phi+{\partial V\over \partial\phi}=0\;,
\end{equation}
with dissipation coefficient
\begin{equation}
\Upsilon=-\int    d^4x'    \,\Sigma_R(x')\,t'=    {i\over    2}\left.{\partial
  \Sigma_\rho(0,\omega)\over\partial \omega}\right|_{\omega=0}\;,
\label{cf}
\end{equation}
where $\Sigma_\rho$ is related to $\Sigma_R$ in Eq. (\ref{srsr}). This equation forms the starting
point for calculating the friction coefficient in particular particle models. The general
expression relates the friction coefficient to the imaginary part of the self energy
$\Sigma_\rho(p,\omega)$ at zero momentum.

\subsection{Coupled field systems}
\label{cfs}

We shall now obtain the friction term in a basic example with an inflaton
field, another scalar field $\chi$, and thermal radiation fields $\sigma$. The
field $\chi$ in this example acts as the only channel for the transfer of
energy from the inflaton into heat radiation. This situation offers the best
prospect so far for realising warm inflation in realistic models, as it gives
some degree of separation between the processes which govern the
thermalisation of the heat bath and the coupling constants which are relevant
to the dissipative dynamics of the scalar field. Model building will be
discussed further in Sec. \ref{particlemod}.

A suitable Lagrangian for the inflaton and $\chi$ interactions is

\begin{equation}
{\cal L}_I=g\,m\,(\delta\phi+\delta\phi^*)|\chi|^2+2g^2|\delta\phi|^2|\chi|^2,
\label{LI1}
\end{equation}
where we use $\phi$ to denote the background value of the inflaton field and
$\delta\phi$ to denote the fluctuating components.  {}For the $\chi$ and $\sigma$
interactions,

\begin{equation}
{\cal L}^\prime_I={1\over \sqrt{2}}h\,m\,(\sigma^2\chi^*+\sigma^{*2}\chi)\;.
\label{LI2}
\end{equation}
Complex fields are used because they embed more easily into supersymmetric
theories, which, as discussed previously in Sec. \ref{wipicture}, are more
suitable to describe realistic warm inflation models by keeping quantum (and
thermal) corrections to the potential, that would be otherwise harmful, small
enough.

\begin{figure}[ht]
  \vspace{1cm} \epsfysize=3cm {\centerline{\epsfbox{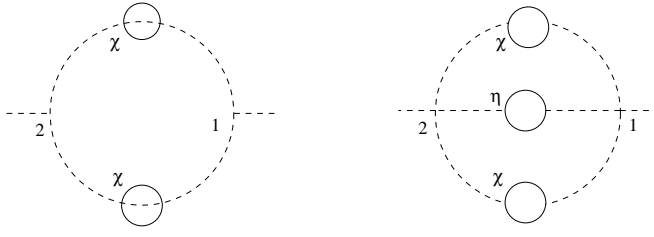}}}
\caption{Contributions  to the $\phi$  self-energy of  order $g^2$  (left) and
  $g^4$ (right).}
\label{simple}
\end{figure}

The contribution to the self-energy of the inflaton field at order $g^2$ is
given by the first diagram in {}Fig. \ref{simple} with two $\chi$ propagators.
When the self-energy is expressed in terms of the spectral function
$\rho_\chi$ of the $\chi$ field, one obtains a formula for the dissipation
coefficient,

\begin{equation}
\Upsilon=4g^2m^2\,   \int   {d^3k\over  (2\pi)^3}\int_0^\infty   {d\omega\over
  2\pi}\, \rho_\chi^2\,n'.
\label{simplegam}
\end{equation}
{}For small $h$ and fixed $T$, the energy integral is dominated by the point
$\omega_k=(k^2+m_\chi^2)^{1/2}$, which lies close to two poles in the spectral
function. These two poles are at $\omega=\omega_k\pm i\tau_\chi^{-1}$, where
$\tau_\chi$ is the relaxation time for the $\chi$ boson defined by Eq.
(\ref{relax}).  The integrand can be expanded about $\omega=\omega_k$ to
obtain a formula first obtained by Hosoya and Sakagami \cite{hosoya1},

\begin{equation}
\Upsilon\approx    g^2m^2\,\beta\int     {d^3k\over    (2\pi)^3}    {\tau_\chi
  \over\omega_k^2}n(n+1).
\label{hs}
\end{equation}
The dependence on a relaxation timescale is typical of the structure which one
might expect from elementary transport theory. Note that reducing the coupling
constant $h$ increases the relaxation time (since $\tau_\chi \sim {\cal
  O}(h^{-2})$) and therefore increases the friction coefficient. This is also
typical of elementary transport theory and is seen, for example, in the Drude
theory of conductivity. The obvious `reductio ad absurdam' argument of
reducing the coupling to zero does not apply because our assumption that the
system remains near thermal equilibrium sets an upper limit to the relaxation
time.

The contribution to $\tau_\chi$ due to the single $\sigma$ loop with coupling
$h$ gives $\omega\tau_\chi^{-1}=h^2m^2/(32\pi)$, 
and \cite{hosoya1,GR,BGR}

\begin{equation}
\Upsilon\approx  {16\over \pi}{g^2\over  h^2}\,T\ln{T\over  m_\chi}\;,\qquad T
\gg m_\chi.
\end{equation}
Note that including other interactions which reduce $\tau_\chi$ will also
reduce $\Upsilon$.

\begin{center}
\begin{figure}[ht]
  \scalebox{1.0}{\includegraphics{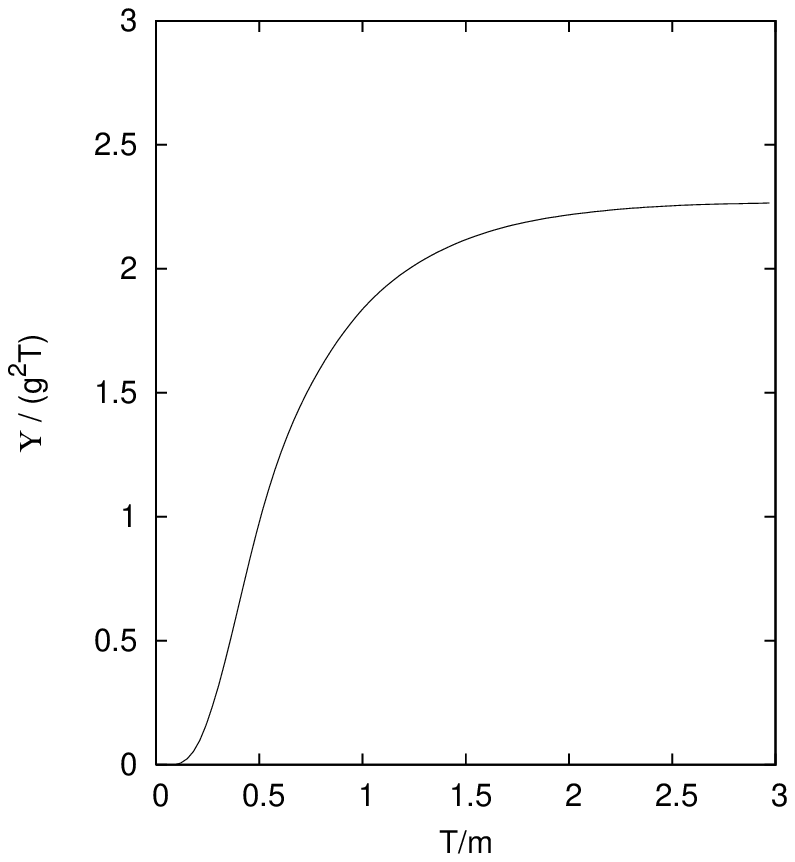}\includegraphics{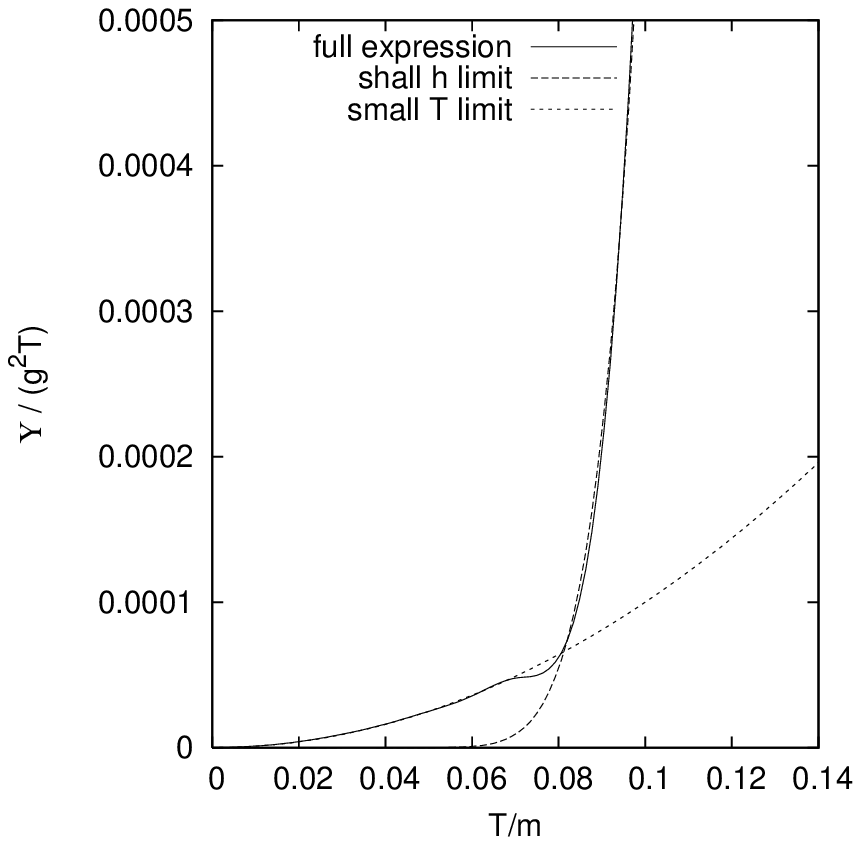}}
\caption{The different  approximations for  the
  friction coefficient are shown in the intermediate and low temperature
  region. The full expression, plotted on the left, corresponds to Eq.
  (\ref{simplegam}) and the low temperature approximation to Eq. (\ref{flt}).
  These plots include both $\phi\to\chi\to2\sigma$ and $\phi\to\sigma\chi$
  decay channels. Coupling constants are $h^2/8\pi=0.025$ and $m_\chi=m$.}
\label{uvt}
\end{figure}
\end{center}

The approximation used to derive Eq. (\ref{hs}) fails at low temperatures,
when the low energy and momentum behavior of the spectral function becomes the
crucial consideration.  This case was first addressed correctly in \cite{mx}.
In cases where the $\chi$-field self-energy is non-vanishing in a neighborhood
of $k=\omega=0$, we use the relation $\rho_\chi\approx
(\Sigma_\rho)_\chi/m_\chi^8$ to deduce that

\begin{equation}
\Upsilon\approx       C       g^2h^4\left(m\over      m_\chi\right)^6{T^3\over
  m_\chi^2}\;,\qquad T \ll m_\chi,
\label{flt}
\end{equation}
for a constant $C$, which can be determined accurately from numerical
integration. {}For the interaction Lagrangian given above, $C\approx 0.006$.
In figure \ref{uvt}, we plot the
overall behaviour for the dissipation coefficient for
different regimes of temperature.

A fully supersymmetric theory has many other interaction terms which
contribute to the dissipation term in the inflaton equation of motion. At high
temperatures we must consider an interaction term,

\begin{equation}
{\cal L}_I =2h^2|\chi|^2|\sigma|^2,
\end{equation}
in addition to those in Eqs. (\ref{LI1}) and (\ref{LI2}).  This term dominates
the $\chi$-field self-energy at large temperatures leading to
$\omega\tau_\chi^{-1}\sim h^4 T^2/(128\pi^2)$ and, from an analogous
expression to the one given by Eq. (\ref{hs}), it gives for the dissipation
coefficient the result \cite{hosoya1,GR,BGR}

\begin{equation}
\Upsilon\sim {64\over\pi}{g^2 \over h^4} m^2T^{-1}\ln{T\over m_\chi}\;, \qquad
hT \gg m_\chi.
\end{equation}

Another interaction which may be present in the supersymmetric theory is

\begin{equation}
{\cal L}_I =gh(\delta\phi\,\chi+\delta\phi^*\,\chi^*)|\sigma|^2.
\end{equation}
This allows a direct interaction between the inflaton and the heat bath, but
it does so without affecting the inflaton potential in any serious way. The
interaction results in a vertex correction to the inflaton self-energy which
modifies the friction coefficient $\Upsilon$, increasing the value of the
constant $C$, in Eq. (\ref{flt}), to $C\approx 0.023$.

{}Fermionic decays can also be included, replacing the scalar field $\chi$ by
a fermion or replacing the scalar field $\sigma$ by a fermion.  These are also
considered in \cite{mx}, where the different dissipation coefficients for each
case can be found.  The fermionic heat bath fields reduce the high temperature
friction coefficient, but at low temperatures they contribute terms $\propto T^5$
to the friction coefficient, which are negligible compared to the bosonic contributions.

\subsection{Physical picture of particle production }
\label{parprod}

So far we have concentrated on the dissipative effects produced by the
interaction between the inflaton and the radiation fields.  The underlying
process here is some form of particle production, and it should be possible to
see the same physics by a consideration of the particle production rates. This
is in fact the case, and this approach offers a way of understanding the
thermalisation processes in the radiation, or indeed, of discussing what
happens when thermalisation is incomplete.

Energy conservation implies that the evolution of the total radiation energy
density is given by

\begin{equation}
\dot\rho_r=\Upsilon\dot\phi^2-4H\rho_r.
\label{ed}
\end{equation}
This equation may have an equilibrium point where the redshift and the
particle production vanish.  In the thermal case, this equation has a stable
equilibrium when $\Upsilon\propto T^3$, but not if $\Upsilon\propto T^4$ or
any higher power of the temperature.  It was therefore very important that the
friction coefficients given in the previous subsection had the necessary $T^3$
behaviour.

We can weaken the thermal assumption and assume a quasiparticle approximation
in which the propagators have a similar form to thermal propagators but where
the momentum distribution function of the radiation fields $n({\bf p},t)$ is
arbitrary.  The energy density is then

\begin{equation}
\rho_r=\int{d^3 p\over (2\pi)^3}n({\bf p},t)\,\omega_p.
\end{equation}

The distribution function for the radiation evolves by a Boltzmann-type of
equation with a source term representing particle production ${\cal S}_P$ from
the evolving inflaton fields, a collision term ${\cal S}_C$ due to the field
interactions and a redshift term ${\cal S}_R$ caused by the expansion of the
universe,

\begin{equation}
\dot n_\sigma={\cal S}_P+{\cal S}_R+{\cal S}_C\;.
\end{equation}
Just as we discussed for the total energy density, equilibrium may occur when
the redshift and the particle production balance, but now we also need the
collision integral to drive the momentum dependence towards a thermal
spectrum.  Given a sufficiently large self-coupling for the radiation fields
there is no reason in principle why thermalisation cannot occur, and numerical solutions
support this conclusion \cite{grahammoss} . An interesting possibility is that departures from a 
thermal distribution can be studied using this approach and their effects on the density
fluctuations analysed. This problem is also well-suited to numerical analysis.

\begin{center}
\begin{figure}[ht]
  \scalebox{1.0}{\includegraphics{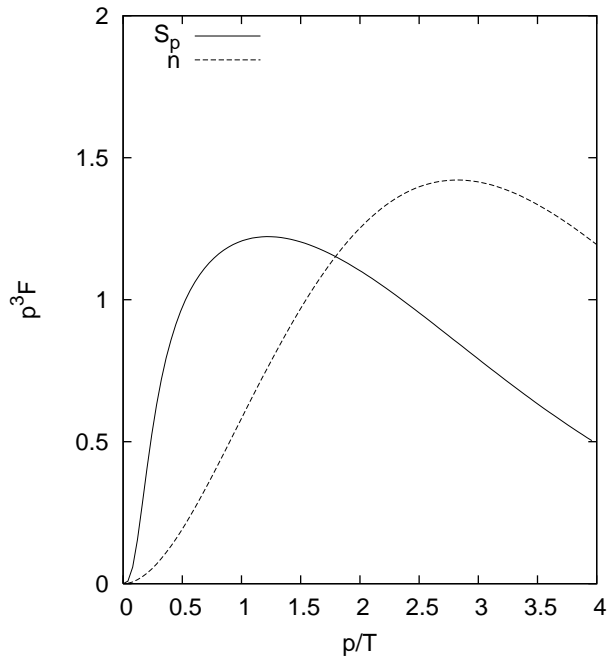}}
\caption{This plot shows the momentum dependence of the particle production rate $S_p$ for the
  production of low mass fields through an intermediate heavy field. The
  thermal distribution $n$ is shown for comparison.}
\label{Sp}
\end{figure}
\end{center}

The source term depends on the details of the particle production mechanism.
{}For the two-stage decay mechanism used in Sec. \ref{cfs}, the slowly
evolving inflaton field cannot produce very massive $\chi$ particles directly,
but it can decay into massless radiation fields via an intermediate virtual
$\chi$ channel.  The source term for massless radiation can be found
analytically \cite{grahammoss},

\begin{equation}
{\cal  S}_P={1\over  256\pi^3}g^2h^4\left({m\over  m_\chi}\right)^6  {T^3\over
  m_\chi^2}   F(p)\,\dot\phi^2\;,  \qquad
T\ll m_\chi\;,
\end{equation}
where $F(p)$ is plotted in figure \ref{Sp}. The momentum distribution is
larger at low momentum than a thermal distribution with similar total energy.
Integrating the Boltzmann equation over momentum recovers the energy density
equation (\ref{ed}) with the same friction coefficient as was obtained before
in Eq. (\ref{flt}). This provides an important consistency check between the dissipation
 and the particle production mechanisms.

\section{Extending dissipative dynamics to 
  curved Friedmann-Robertson-Walker space-time }
\label{FRWspacetime}

In order to complete our discussion of dissipative dynamics we now consider
how the results quoted so far extend to curved space-time. Specifically, we
consider a homogeneous and isotropic, spatially flat,
Friedmann-Robertson-Walker metric with scale factor $a$. The naive expectation
would be that thermal effects are more important than curved space quantum
effects when we have radiation with temperature $T$ which is much larger than
the expansion rate $H$, $T\gg H$.  This expectation can be supported by an
approximation scheme given below, and therefore curved space quantum effects
can be considered small during warm inflation, at least during a thermal
regime.

Scalar field propagators can be defined as in Eq.  (\ref{propdef}) on the
{}Friedmann-Robertson-Walker background and satisfy a curved space version of
the propagator equation,

\begin{eqnarray}
\left[\partial_t^2 +  3 H\partial_t -a^{-2}\nabla^2+ m^2 +  \zeta R(t) \right]
G_{ab}  (x,x') + \int  d^4 y\,  \Sigma_{a}{}^c (x,y)  G_{cb} (y,x')  = ic_{ab}
\delta (x,x') \;,
\label{Gchifrw}
\end{eqnarray}
where $R$ is the curvature scalar, $R=6\dot H+12H^2$, and $\zeta$ is
dimensionless parameter describing the coupling of matter fields to the
gravitational background. Both $d^4y$ and $\delta(x,x')$ implicitly include
$\sqrt{-g}$ factors, where $g$ is the metric determinant, e.g.

\begin{equation}
\delta(x,x')={\delta^4(x-x')\over a^{3/2}(t)a^{3/2}(t')},
\end{equation}
as required by general covariance.

The Lagrangian density for the inflaton field given in Eq. (\ref{S12}) is now,

\begin{equation} 
{\cal L} [ \phi] =  \sqrt{-g} \left( \frac{1}{2} g^{\mu \nu} \partial_\mu \phi
\partial_\nu  \phi -  \frac{m_\phi^2}{2}\phi^2 -  \frac{\lambda}{4  !}  \phi^4
-\frac{\zeta_\phi}{2} R \phi^2 \right)\;.
\label{Lfrw} 
\end{equation} 

We can decompose the propagator and self-energy matrices as we did in flat
space. Thus, for the Lagrangian density model given by Eq. (\ref{Lfrw}), it
follows by the same arguments given in Sec. \ref{effeom} that the effective
stochastic equation of motion for the background scalar field $\phi_c$ is
given by

\begin{equation}
\left[\partial_t^2 + 3H\partial_t -a^{-2}\nabla^2+ m_{\phi}^2 + 
\zeta_\phi R(t)+
     {\lambda\over       3!}\phi_c(x)^2\right]\phi_c(x)       +\int      d^4x'
     \Sigma_R(x,x')\phi_c(x') = {\xi(x)\over a^3},
\label{langevinphiRW}
\end{equation}
where noise field $\xi(x)$ again satisfies $\langle \xi(x) \rangle=0$ and it
has the same two-point function as in Eq. (\ref{noisetwopoint}), but where the
self-energy terms are expressed in terms of propagators in curved space-time.

These expressions hide the fact that the propagators for the fields in curved
space-time are in general very complicated functions \cite{semeno,mallik2}.
{}For the situation of interest here, we have a homogeneous background with
scalar and radiation fields present. In the spatially homogeneous case, let

\begin{equation}
G_{ab}({\bf  k},t,t')=\int  d^3x\,G_{ab}(x,x')\,e^{i{\bf k}\cdot({\bf  x}-{\bf
    x'})}\;.
\end{equation}
We shall make a pseudoparticle (or Kadanoff-Baym) approximation for the
radiation (and $\chi$ field) propagators, introducing an occupation number
$n({\bf k},t)$. This may, for example, be thermal at some initial time $t_0$
with frequency $\omega(t_0)$, where

\begin{equation}
\omega(t)^2 = k^2/a^2 + M^2,
\end{equation}
with $M$ the quadratic mass term for the field in the curved space-time. The
spectral and anticommutator functions defined by Eqs. (\ref{rho}) and
(\ref{functionF}), respectively, in the pseudo-particle approximation are

\begin{eqnarray}
\rho({\bf   k},t,t')&=&i\left[   f_1({\bf   k},t)f_2({\bf   k},t')-   f_2({\bf
    k},t)f_1({\bf      k},t')\right]\\      F({\bf     k},t,t')&=&\left[n({\bf
    k},t)+\frac12\right]   \left[   f_1({\bf   k},t)f_2({\bf   k},t')+f_2({\bf
    k},t)f_1({\bf k},t')\right]\;,
\label{Fkb}
\end{eqnarray}
where the functions $f_{1,2}({\bf k},t)$ are defined by the solutions of the
differential equation \cite{semeno2}

\begin{eqnarray}
\left[  \frac{d^2}{dt^2} + 3  \frac{\dot{a}}{a} \frac{d}{d  t} +
  \frac{k^2}{a^2} + M^2 (t) \right] f_{1,2}({\bf k},t) =0 \;,
\label{modesf}
\end{eqnarray}
with Wronskian $\dot{f}_1(t)f_2(t)-f_1(t)\dot{f}_2(t)=-i/a^3(t)$.

Usually, solutions for Eq.  (\ref{modesf}) are known only for some specific
cases, e.g.  for de Sitter expansion $H\sim$ constant, so $a(t) = \exp(H t)$
\cite{davis}, and power law expansion $a(t) \sim t^n$ \cite{habib}.  {}For
example, in de Sitter the solutions for constant $M$ are given in terms of
Hankel functions,

\begin{equation}
f_{1} ({\bf k},t)=  f_{2}^* ({\bf k},t)= {\sqrt{\pi}\over 2}H^{-1/2}e^{-3Ht/2}
H^{(1)}_{\nu} \left(k e^{-Ht}/H \right)\;,
\label{exp_modes}
\end{equation}

\noindent
with $\nu^2 =9/4-M^2/H^2 - 12 \zeta_{\phi}$.

Alternatively, adiabatic approximations for the mode functions can be derived
by applying a WKB approximation to Eq. (\ref{modesf}) \cite{mallik2,semeno2},
\begin{equation}
f_{1}({\bf   k},t)=f_{2}^*({\bf  k},t)  \approx   \frac{1}{a^{3/2}  (t)\sqrt{2
    \omega(t)} } \exp\left[- i\int^t_{t_0} d t' \omega(t') \right]\;.
\label{F mode}
\end{equation}
Note that this leads to a well defined split between positive and negative
frequency modes.

The WKB approximation requires the adiabatic conditions $\omega^2\gg
\dot\omega^2/\omega^2,\ \ddot\omega/\omega$, which apply at large mass or
large momentum.  Both of these are relevant to models of warm inflation, where
we may have large mass (for the $\chi$ field) and large momentum due to large
temperature (compared to the expansion rate), as shown in Sec.
\ref{particlemod}.

These adiabatic conditions also simplify the equations for the propagators and
allow interactions to be taken into account. Berera and Ramos in \cite{BRfrw}
have demonstrated both numerically and also analytically that the dynamics of
the inflaton including dissipation effects, is well approximated by the
Minskowkian dynamics, thus allowing us to use the results in Subsec. \ref{cfs}
for the dissipation coefficients in warm inflation.

The energy density of the radiation $\rho_r$ in the pseudo-particle
approximation can be expressed in terms of the anti-commutator function
(\ref{Fkb}).  {}For massless radiation in de Sitter space, the energy density
is given by

\begin{equation}
\rho_r=\rho_{\rm   de\   Sitter}+{1\over   a^4}\int  {d^3k\over   (2\pi   )^3}
\left(1+{1\over 2}{a^2\over H^2k^2}\right)\,k\, n(k,t)\;,
\end{equation}
where $\rho_{\rm de\ Sitter}$ is the energy density of the de Sitter vacuum. A
consequence of this expression is that the energy density rapidly redshifts
towards the de Sitter value if $n(k,t)$ is time independent. On the other
hand, the combination of particle production and particle interactions can
produce a thermal distribution in the physical energy spectrum as in Sec.
\ref{parprod}, {\it i.e.}, $n\equiv n(k/a)$ and then we recover a thermal
contribution to the energy density.

\section{Particle Physics Model Building for Warm Inflation}
\label{particlemod}

Several phenomenological warm inflation models have been constructed
\cite{Berera:1996fm,dr,gmn,mb,lf,ml,Billyard:2000bh,tb,Donoghue:2000fk,dj,Dymnikova:2001jy,cjzp,DeOliveira:2002wk,hmb,Jeannerot:2005mc,Mimoso:2005bv,ch,Hall:2007qw,Battefeld:2008py,Mohanty:2008ab}.  
Moreover, interesting
applications of the warm inflation regime have been suggested for generating
cosmic magnetic fields \cite{Berera:1998hv} and for baryogenesis \cite{by}.
More interesting are models of warm inflation constructed completely from
first principles quantum field theory.  There are by now many such models and
these models have some unique and attractive features.{} First, in the strong
dissipative regime they have no ``$\eta$"-problem.  
The ``$\eta$"-problem typically
emerges since supergravity (SUGRA) corrections to the inflaton mass are of
order the Hubble scale, yet in cold inflation the inflaton needs a mass less
than the Hubble scale to realise slow-roll inflation
\cite{Copeland:1994vg,Dine,Gaillard:1995az,Kolda:1998kc,Arkani-Hamed:2003mz}.  
In contrast, in strong dissipative warm inflation, the inflaton mass is much
bigger than the Hubble scale, so such models are fairly insensitive to SUGRA
corrections.  The second attractive feature of warm inflation models is for
monomial potentials the amplitude $\phi$ of the inflaton field is always below
the Planck scale.  In cold inflation, for monomial potentials \cite{ci}, the inflaton amplitude
during inflation is larger than the Planck scale.  This is a problem for model
building.  Quantum field theory models are generally regarded as low energy
effective theories of some higher more fundamental theory, such as possibly
strings, and there is some upper energy scale to which this low energy
approximation is valid.  Above this scale, the theory would be modified by
additional, usually an infinite number of, operator corrections.  The highest
scale can be the Planck scale, and so for any quantum theory above this scale
one expects an infinite number of nonrenormalisable operator corrections,
$\sim \sum_{n=1}^{\infty} g_n \phi^{4} (\phi/m_{P})^n$, which have to be
retained
\cite{Copeland:1994vg,Gaillard:1995az,Kolda:1998kc,Arkani-Hamed:2003mz}.  In
such a regime the low energy approximation to the theory is essentially not
useful.  Thus from a model building perspective cold monomial inflation models
are difficult to implement, whereas warm inflation models discussed in this
section do not suffer this complication.

Hilltop inflation models \cite{German:1999gi,Boubekeur:2005zm}
are one type of single field cold inflation
model where the inflaton amplitude remains below the Planck scale
initially during inflation, thus when the large scale features are
determined. However even in these models typically the inflaton
amplitude goes above the Planck scale by the end of inflation.
Beyond these models, maintaining the inflaton amplitude below
the Planck scale in cold inflation models
requires more elaborate constructions, with a few
examples such as \cite{Ross:1995dq,Adams:1996yd,Lyth:1998xn,Burgess:2007pz}.
However for warm inflation, as will be seen in this section
the simplest monomial models are consistent both
from the cosmological and particle physics perspective.

One generic feature of warm inflation models is that they require a large
number of fields, usually in the hundreds or more. {}From the perspective of
the simple single or few field inflation models typically seen in cold
inflation, this feature of warm inflation models may appear undesirable.
However in most high energy particle physics models, there are typically many
fields present.  In fact, from the perspective of string theory, where there
are a huge number of fields, warm inflation models can look very compelling.
Thus, in its own right, the requirement of a large number of fields
distinguishes warm inflation models from their cold counterparts, but one can
not say that they are necessarily more peculiar.

To calculate a warm inflation model it requires solving the evolution equation
(\ref{wieom}) and constraining the result with the density fluctuation
amplitude and spectral index.  In addition, for a first principles model, the
dissipative coefficient $\Upsilon$ has been computed from an underlying
quantum field theory model.  One assumption in such calculations is that the
microphysical dynamics determining $\Upsilon$ is operating at time scales much
faster than that of the macroscopic motion of the inflaton background field
and the expansion scale of the Universe.  To realise such an adiabatic regime
leads to a set of consistency conditions \cite{BGR},

\begin{equation}
\tau_i^{-1} > {\dot \phi}/ \phi, H ,
\end{equation}
where $\tau_i^{-1}$ here represents all relevant decay widths of the fields
responsible for dissipation.

Another challenge in realising warm inflation models from first principles is
the effect of radiative and thermal corrections to the effective potential,
which if too large would ruin the necessary flatness of the inflaton
potential.  There are competing requirements in that large dissipation prefers
large couplings whereas controlling radiative and thermal corrections requires
small couplings.  Supersymmetry provides a means to achieve both these
requirements.  The observation is that SUSY will cancel local radiative
corrections such as the loop corrections to the effective potential. Thus the
use of SUSY models with interactions like the ones shown in Sec. \ref{fd}.
However SUSY is ineffective in cancelling time nonlocal loop effects from which
dissipation emerges.  Of course there are still limitations. {}For one thing,
since inflation requires a nonzero vacuum energy, SUSY must be broken, thus
cancellations of radiative corrections are never perfect.  Moreover, SUSY is
also broken at finite temperature and so thermal loop corrections do not
cancel exactly, although a significant amount of cancellation does occur
\cite{Hall:2004zr}.

With these considerations in mind, several warm inflation models have by now
been constructed.  The first of these were the distributed mass (DM) models
\cite{BGR2}.  In these models there are a set of bosonic fields $\chi_{i}$
coupled to the inflaton field through shifted couplings.
The interaction
term in the Lagrangian which realizes such shifted couplings has the form,
\begin{equation}
\frac{g^2}{2} (\phi - M_i)^2 \chi_i^2 ,
\label{dmmchiint}
\end{equation}
so that when $\phi \sim M_i$, the mass of the $\chi_i$ field becomes small.  In
particular the mass is meant to get below the temperature scale in the
Universe, so that these $\chi$ fields can become thermally excited.  Once
thermally excited, as the background inflaton field evolves, it is able to
dissipate energy into these fields.  This results in a dissipative term in the
inflaton evolution equation \cite{BGR}.  If these mass scales $M_i$ are now
distributed over a range of values that $\phi$ will go through, then during
evolution of $\phi$, some subset of these $\chi$ fields will be light and
generate a dissipative term.  In order to control the radiative corrections in
this model, SUSY has to be implemented. A simple superpotential that realises
this model is
\begin{equation}
W =  4m\Phi^2 + \lambda \Phi^3  + \sum_{i=1}^{N_M} \left[2gM_i  X_i^2 + fX_i^3
  -2g \Phi X_i^2 \right] .
\label{dmmsp}
\end{equation}
Here the bosonic part of the chiral superfield 
$\Phi = \phi + \theta \psi + \theta^2 F$, with
$\theta \psi \equiv \theta^{\alpha} \psi_{\alpha}$ and 
$\theta \equiv \theta^{\alpha} \theta_{\alpha}$, is the
inflaton field $\phi$, 
and it interacts with both the bose and fermi fields of the
chiral superfields $X_i = \chi_i + \theta \psi_{\chi_i} + \theta^2 F_{\chi_i}$.
The potential terms of the Lagrangian are obtained from Eq. (\ref{dmmsp})
by standard procedures;  the potential is
$L_V = \int d^4x d^2 \theta W(\Phi,\{X_i\}) + h.c.$
and the auxiliary fields $F$ and $F_{\chi}$ are eliminated through the
``field equations'', 
$\partial W/\partial F = \partial W/\partial F_{\chi_i} = 0$,
which results in the Lagrangian only in terms of the
bose and fermi fields.
This leads for the above superpotential Eq. (\ref{dmmsp})
to a $\lambda \phi^4$ inflaton potential with
interactions to the $\chi_i$ fields similar to Eq.  (\ref{dmmchiint}) and
corresponding interaction terms to the $\psi_{\chi_i}$ fermi fields.  The mass
scales $M_i$ are distributed along the line which $\phi$ traverses during the
inflationary period.
Moreover the $\phi^4$ coupling must remain small
for successful inflation, and in this SUSY theory
it occurs since the renormalization group equations
for the quartic coupling are proportional to the coupling
itself, thus even if there is another large coupling
this will not lead to a problem.
This model can generate warm inflation with adequate
e-foldings to solve the horizon and flatness problems \cite{BGR2} as well as
produce observationally consistent primordial fluctuations
\cite{Berera:1999ws}.
It should be noted that the most general superpotential would
include a term in Eq. (\ref{dmmsp}) linear in the $X_i$ fields,
$\Phi^2 X_i$ and this term has been eliminated by hand.
This term induces a $\phi$ dependent
mass term to all the $X_i$ fields and so
must be very small for the success of this model.
The stability of the SUSY theory under radiative corrections
allows this term to be eliminated by hand. However one
can prohibit the linear term in the superpotential
more elegantly by imposing a charge under some,
for example GUT, symmetry so that these $X_i$ fields are not
singlets.

In \cite{Berera:1998cq,Berera:1999wt} the DM model has been shown as arising
from a fine structure splitting of a {\it single} highly degenerate mass
level.  {}For typical cases studied in \cite{BGR2,Berera:1999ws}, it was shown
in \cite{Berera:1999wt} that for significant expansion e-folding, $N_e>60$, if
$M \approx g|M_{i+1} - M_i|$ denotes the characteristic splitting scale
between adjacent levels, warm inflation occurred in the interval $10^3M
\stackrel{<}{\sim} \phi \stackrel{<}{\sim} 3 \times 10^3M$ and of note, at
temperature $M \stackrel{<}{\sim} T$ and not $T$ at the much higher scale of
the mass levels $\sim 10^3M$.  What makes these massive states light is
precisely the shifted mass couplings.  In the string picture, this arrangement
corresponds to a fine structure splitting of a highly degenerate state of very
large mass, $\sim M_S$, with the fine structure splitting scale several orders
of magnitude less than the mass of the state, say $M \stackrel{<}{\sim}
M_{GUT} \sim 10^{-3}M_S$.

In Ref. \cite{Berera:1999wt} the following string scenario was suggested.
Initially in the high temperature region, some highly degenerate and very
massive level assumes a shifted mass coupling to $\phi$.  Since all the states
in this level are degenerate, at this point they all couple identically as
$g^2 \sum_i (\phi-M)^2 \chi_i^2$.  The string then undergoes a series of
symmetry breakings that split the degeneracy and arrange the states into a DM
model $\sum_i (\phi-M_i)^2 \chi_i^2$ with $0 < (M_i-M_{i+1})/M_i \ll 1$.

This string scenario has several appealing features:

\begin{itemize}
\item[(i)] Strings have an ample supply of highly degenerate massive states.
\item[(ii)] The generic circumstance is that as temperature decreases, many of
  the degeneracies will break at least a little, and for warm inflation a
  little is all that is needed. Moreover, warm inflation occurs when $T$ is at
  or above the fine structure splitting scale but much below the scale of the
  string mass level.  Thus, for the respective mass level, warm inflation is
  occurring in a low temperature region. This further supports the expectation
  that degeneracies for that level have broken.
\item[(iii)] The shifted mass coupling to $\phi$ is much more probable to
  occur to a single mass level, albeit highly degenerate, as opposed to the
  coincidence probability to several mass levels.
\item[(iv)] There are minimal symmetry requirements for interactions.  Since
  zero modes and any higher mass level modes fall into representations of the
  gauge and Lorentz groups, the interacting fields must tensor together to
  form singlets.
\end{itemize}

The distributed mass models are the only warm inflation models constructed in
which the fields directly interacting with the inflaton are thermally excited.
In general it is too difficult to control adequately the thermal loop
corrections to the inflaton effective potential to maintain the needed
flatness of the potential.  This has led to the development of the two stage
dissipative mechanism of warm inflation \cite{BR1} in which the inflaton
$\phi$ is coupled to a set of heavy fields $\chi$ and $\psi_{\chi}$ which in
turn are coupled to light fields $y$ and $\psi_{y}$.  The main point is the
heavy fields are not thermally excited and so the loop corrections to the
inflaton potential are only from vacuum fluctuations, which can be controlled
by SUSY.  A generic superpotential that realises the two stage mechanism is
\begin{equation}
W_I = \sum_{i=1}^{N_{\chi}} \sum_{j=1}^{N_{\rm decay}} \left[g \Phi X_i^2 + 4m
  X_i^2 + hX_iY_j^2 \right],
\label{wi2stage}
\end{equation}
where $\Phi = \phi + \psi \theta + \theta^2 F$, $X = \chi+ \theta \psi_{\chi}
+ \theta^2 F_{\chi}$ and 
$Y = \sigma + \theta \psi_{\sigma} + \theta^2 F_{\sigma}$
are chiral superfields.  The field $\phi$ will be identified as the inflaton
in this model with $\phi = \phi_c +\eta$ and $\langle \phi \rangle =
\phi_c$.  In the context of the two stage mechanism $X$ is the heavy fields
to which the inflaton is directly coupled and these fields in term are coupled
to light $Y$ fields.  In order to see the effect of this interaction structure
on radiative corrections to the inflaton potential, a particular model has to
be chosen.  Thus, considering the case of a monomial inflaton potential and
adding the superpotential term $W_{\phi} = \sqrt{\lambda} \Phi^3 / 3$ so that
\begin{equation}
W = W_{\phi} + W_I ,
\end{equation}
this model generates at tree-level the inflaton potential
\begin{equation}
V_0(\phi_c) = \frac{\lambda}{4} \phi_c^4 .
\label{vophi}
\end{equation}
When $\phi_c \ne 0$, there is a nonzero vacuum energy and so SUSY is broken.
This manifests in the splitting of masses between the $\chi$ and $\psi_{\chi}$
SUSY partners with in particular

\begin{eqnarray}
m_{\psi_{\chi}}^2 & = & \left[ 2  g^2 \phi_c^2 + 16\sqrt{2}mg \phi_c + 64m^2
  \right]\;,  \nonumber  \\  
m_{\chi_1}^2  &   =  &  \left[  \frac{1}{8}  (g^2  +
  \frac{1}{2}\sqrt{\lambda} g) \phi_c^2 + \sqrt{2}mg \phi_c + 4m^2 \right] =
m^2_{\psi_{\chi}} + \sqrt{\lambda}g \phi_c^2\;,  \nonumber \\ 
m_{\chi_2}^2 & = &
\left[ \frac{1}{8}  (g^2 - \frac{1}{2}\sqrt{\lambda}  g)\phi_c^2 + \sqrt{2}mg
  \phi_c + 4m^2 \right] = m^2_{\psi_{\chi}} - \sqrt{\lambda}g \phi_c^2 .
\label{cpmass}
\end{eqnarray}
The one loop zero temperature effective potential correction in this case is
\begin{equation}
V_1(\phi_c)    \approx   \frac{9}{128    \pi^2}    \lambda   g^2    \phi_c^4
\left(\ln\frac{m^2_{\psi_{\chi}}}{m^2}   -2  \right) \ll   V_0(\phi_c)  =
\frac{\lambda}{4} \phi_c^4\;,
\end{equation}
which is further suppressed than the tree level potential Eq. (\ref{vophi})
and so will not alter the flatness of the inflaton potential.  There are
several first principles warm inflation models that implement the two-stage
mechanism Eq. (\ref{wi2stage}), which will be summarized here.

\subsection{monomial potential}

The general form of the inflaton monomial potential to be studied is,
\begin{equation}
V(\phi) = V_0 \left(\frac{\phi}{m_P}\right)^n \label{vpoly}\,,
\end{equation}
with $n>0$. Without enough dissipation, {\it i.e.}, either for cold inflation
with $Q=0$, or only weak dissipation with $Q < 1$, 
where $Q$ is defined in Eq. (\ref{uhratio}),
these kind of models lead
to inflation only for values of the inflaton field larger than the Planck mass
$m_P$. On the other hand, in the strong dissipative regime due to the larger
friction term, slow-roll conditions Eqs. (\ref{slowwi}) can be fulfilled for
values of the field well below the Planck scale. Thus Eq. (\ref{vpoly}) can be
regarded from the effective field theory point of view, with the potential
well defined below the cut-off scale $m_P$; higher order term contributions
suppressed by $m_P$ will be then negligible, without the need of fine-tuning
the coefficients in front.  In \cite{bb4} ${\cal N}=N_\chi N^2_{decay}$ and
$g_*$ were treated as free parameters, and it was examined for which values
$\eta/Q$ and $T/\phi$ can be kept small enough for at least 50 e-folds or so
(and $T/H >1$). {}For example, for a quartic potential with $V(0)^{1/4} \simeq
0.3 m_P$ and $\phi(0)= m_P$, in order to satisfy all the constraints it
required $g_* < 100$ but ${\cal N} > 2300$. Similar results were obtained for
other powers of the potential. By lowering the value of the potential, it was
found to be easier to fulfill all conditions except that for the ratio
$\eta/Q$. Keeping the latter below one gives the lower bound:
\begin{equation}
{\cal N} > 8.4  \times 10^{-2}\, \frac{ g_*^{3/4} m_P}{V(0)^{1/4}} 
\left[ n^{1/7} (n -1) + \frac{n}{7} N_e \right]^{7/4} \,,
\end{equation}
and the lower $V(0)$ is, the larger ${\cal N}$ has to be. {}For example, for
$n=4$, $V(0)^{1/4}/m_P = 0.1$, and $g_*= 10$, it requires ${\cal N} > 2800$,
but getting to the number of degrees of freedom for the MSSM, $g_*= 228.75$,
would require ${\cal N} > 29000$.  The interesting result from this analysis
was that due to the extra friction, inflation occurred for values of the field
below the Planck scale, although the model prefers an initial value of the
height of the potential only an order of magnitude or so below the Planck
scale.

On the other hand, the amplitude of the primordial spectrum is also affected
by the strong dissipative friction term and the presence of a thermal bath.
In order to keep the amplitude of the primordial spectrum consistent with
WMAP's value \cite{Komatsu:2008hk}, $P_{\cal R}^{1/2} \simeq 5.5 \times
10^{-5}$, it required a potential much smaller than ${\cal O}(10^{-12}
m^4_P)$. {}For such a value of the potential, it needs roughly ${\cal N} \sim
{\cal O}(10^6)$ in order to get at least 50 e-folds in the strong dissipative
regime.

\subsection{hybrid potential}

In \cite{bb4} small field models of inflation were also considered of the
form,
\begin{eqnarray}
V(\phi) &=& V_0 \left[ 1 + \left(\frac{\phi}{M}\right)^n \right]
\label{vhybrid}\,,\;\;\; n > 0\;, \\
V(\phi) &=& V_0 \left( 1 + \beta \ln\frac{\phi}{M}\right)
\label{vshybrid}\,,\;\;\; n = 0\;.
\end{eqnarray}
Given that during inflation the potential is dominated by the constant term
$V_0$, the value of the field can easily be kept below the Planck scale in
these models during slow-roll inflation.  These potentials can be regarded as
a generalisation of a hybrid model \cite{Linde:1993cn,Copeland:1994vg}, where
inflation ends once the inflaton field reaches the critical value,
destabilising the waterfall field coupled to it. Those interactions are not
relevant to study the slow-roll dynamics, only to mark the end of inflation,
and therefore they do not need to be considered in the inflationary potential
Eqs. (\ref{vhybrid}) and (\ref{vshybrid}).  As observed in \cite{bb4}, the
same interactions between the inflaton and the waterfall field required by the
hybrid mechanism will give rise to dissipation, and leads in the low-T regime
to the dissipative coefficients given in Subsec. \ref{cfs}.  The case $n=2$
would be the standard hybrid model \cite{Linde:1993cn,Copeland:1994vg}, with a
mass term for the inflaton, whereas $n=0$ is the susy model with the
logarithmic correction coming from the one-loop effective potential
\cite{dvali,lazarides}. 

In supersymmetric hybrid models, one needs to worry
about the $\eta$-problem discussed earlier
\cite{Copeland:1994vg,Dine,Gaillard:1995az,Kolda:1998kc,Arkani-Hamed:2003mz}, 
{\it i.e.},
the fact that generically SUGRA corrections give rise to scalar masses of the
order of the Hubble parameter, including that of the inflaton, which in turn
forbids slow-roll inflation.  Different solution to this problem exist in the
literature, for example by combining specific forms of the superpotential and
the K\"ahler potential \cite{Copeland:1994vg,dvali,lazarides,BasteroGil:2006cm}.  
Nevertheless, typically,
although the quadratic correction can be avoided, {\it i.e.}, a mass
contribution, SUGRA corrections manifest as higher powers in the inflaton
field \cite{sugra}. In the case of strong dissipative warm inflation, the
presence of the extra friction term alleviates the problem: slow-roll
conditions are fulfilled also for inflaton masses 
in the range 
$H \stackrel{<}{\sim} m_\phi \stackrel{<}{\sim} \sqrt{H\Upsilon}$.
In addition, the values of the field being smaller than in standard
cold inflation, the effect of higher order SUGRA corrections is also
suppressed.  

It also should be noted that in all the warm inflation models
constructed in this Section, the inflaton field $\Phi$ is
a singlet, for which an extra complication arises in that
a linear term in $\Phi$ is allowed in the K{\"a}hler potential.
This induces a tadpole term for the singlet, which results in a
large vacuum expectation value (VEV) of the order the cutoff scale.
This can lead to problems for both the low-energy
theory in destabilizing the electroweak scale \cite{Bagger:1993ji}
as well as make it difficult to 
realize inflation \cite{Boubekeur:2001ys}.
Slight modifications have been shown can keep the singlet corrections
under control for both the low-energy theory
\cite{Nilles:1997me,Kolda:1998kb}
as well as inflation \cite{Boubekeur:2001ys,Battye:2006pk}.
Without a higher theory, the terms in the K{\"a}hler potential
ultimately are arbitrary and one can always tune the linear
term to be small. Alternatively one can impose a symmetry on
$\Phi$ which prohibits such a term.  As yet, warm inflation
model building has not explored
these detailed questions. However one point should be noted,
that the presence of the friction term in warm inflation allows for larger
$\eta$ and $\epsilon$, and so the effect of all terms including
the linear term in
the K{\"a}hler potential is alleviated.

The amplitude of the primordial spectrum is given by Eq.  (2.28) and the
spectral index $n_S$ is found in \cite{bb4} to be
\begin{eqnarray}
n_S  -1   &\approx&  \frac{3   \eta  }{7  Q}\left[   \frac{7  -  n}{n   -1}  +
\left(\frac{\phi}{m_P}\right)^2      \frac{3      \eta}{2      (n-1)^2}\right]
\,, \label{nsapprox}
\end{eqnarray}
and
\begin{eqnarray}
n_S^\prime       &       \approx&      -3       \left(\frac{\eta}{7Q}\right)^2
\left[\frac{n(7-n)}{(n-1)^2} +\left(\frac{\phi}{m_P}\right)^2 \frac{(14+10 n -
  17 n^2)}{(n-1)^4} \eta\right]
\label{runapprox} \,.
\end{eqnarray}
where it is assumed $\epsilon \ll \eta$.  Notice that the spectral index is of
the order of ${\cal O}(\eta/Q)$, whilst the running is of the order of ${\cal
  O}(\eta^2/Q^2)$.  Therefore, the same condition needed to have slow-roll in
the strong dissipative regime will avoid having too large a spectral index.
The model has a blue-tilted spectrum when $n \leq 7$, including the case of
$n=0$, {\it i.e.}, the logarithmic potential.  The more negative running, the
more blue-tilted the spectrum can be, which would be the case for $ 0 < n < 7$
with $n_S^\prime \approx -(n_S-1)^2 n/(21-3n)$.

Given that the field decreases during inflation, so does $\eta/Q$, and also
$\rho_R/V$ (or equivalently $T/H$) for any power $n \neq 0$, being constant
for the logarithmic potential $n=0$.  Therefore, the energy density in
radiation will never dominate in this regime.  On the other hand, $\phi/T$
diminishes for $n < 4$, but $Q$ only decreases for $n > 2$. Therefore, for a
logarithmic or quadratic potential once the system is brought into the strong
dissipative regime, it stays there until the end. Indeed one can start in the
weak dissipative regime, with $T>H$ but $Q< 1$, and it will evolve into $Q>1$.

{}For $n=0$ the model exhibits the interesting situation of a transition from
``cold'' $\rightarrow$ ``weak warm'' $\rightarrow$ ``strong warm'' inflation.
{}For $n=2$, $T$ remains constant until it diminishes in the strong
dissipative regime, so that the parameter space divides into either cold or
warm inflation, but the system can evolve from weak to strong dissipation. In
the weak dissipative regime, quantum fluctuations of the inflaton field will
also have a thermal origin, with spectral index \cite{bb1}
\begin{equation}
n_S \simeq 1 - 2 \epsilon + 2 \frac{\eta}{n-1} \,,
\end{equation}
where note that when $n=1$, it also follows $\eta=0$ and so there is no
singularity.  Notice that even in the weak dissipative regime a logarithmic
potential gives rise to blue-tilted spectrum, while $n >0$ leads to a
red-tilted spectrum, just the reverse than the standard cold predictions.
{}For larger powers $n>2$, the opposite behavior is found, so even if
inflation starts in the strong dissipative regime it will evolve towards the
weak and the cold regime. When this happens before the last 50 e-folds of
inflation, then dissipation becomes irrelevant.

There are a few specific interesting features of these models found in
\cite{bb4}, which are discussed next for $n=0,2,4$.

\vspace{3mm} {\bf Case $n=0$: Hybrid logarithmic potential.}  This model has
its lower bound on ${\cal N}$ in the limiting case for slow-roll warm
inflation, when $T/H \simeq 1$, $Q(0) \simeq 1$ and $(\phi/T)_{N} \simeq 10$,
which gives:
\begin{equation}
{\cal N} \simeq 11.87 \exp[-0.23 N_e g_*^{1/2}/{\cal N}^{1/2}] \,.
\end{equation}
For example, with $g_*\simeq 10$, $N_e=50$ we have the lower bound: ${\cal
  N}\simeq 180$, $\phi(0)/H \simeq 150$, $a^2 \simeq 244$, $\eta/Q \simeq
-0.054$; with $g_*\simeq 228.75$, $N_e=50$ we have: ${\cal N}\simeq 1350$,
$\phi(0)/H \simeq 1.2 \times 10^3$, $a^2 \simeq 5.2 \times 10^3$, and $\eta/Q
\simeq -0.1$. Those values of ${\cal N}=N_\chi N_{decay}^2$ are quite in the
range of a realistic model; for example with $\chi$ in the {\bf 126} or {\bf
  351} of $SO(10)$ ($E_6$), and $N_{decay}^2 \approx {\cal O}(10)$, one can
expect having ${\cal N}$ in the range of a few thousands. However, for such
values in \cite{bb4}, the amplitude was found to be too large.  In order to
match the amplitude of the primordial spectrum with WMAP's value, it requires
a larger initial value of the field $\phi(0)$, but then the value of $C_\phi$
(${\cal N}$) has to be larger in order to stay within the low $T$
approximation, with $\phi/T \geq 10$.  Satisfying WMAP's constraints requires
then ${\cal N} \gtrsim 5\times 10^5$, which is rather large, with $\phi(0)/H
\gtrsim 2 \times 10^{6}$ and $a^2 \gtrsim 8\times 10^{11}$, where
\begin{equation}
a^2       \equiv      \left\{\begin{array}{ll}      \displaystyle\left(\frac{n
  V_0}{H^4}\right)\left(\frac{H}{M}\right)^n            \,,&n\neq            0
\,,\\ \displaystyle\left(\frac{\beta  V_0}{H^4}\right) \,,&n= 0 \,.\end{array}
\right.
\end{equation}

An alternative example in \cite{bb4} was inflation in the weak warm regime,
and a transition from weak to strong dissipation at the end.  This helps in
fixing the amplitude of the spectrum to lower values. The following conditions
were still imposed, that (a) $T/H> 1$ (b) to obtain enough inflation, {\it
  i.e.}, $N_e \approx 50$. These translated into ${\cal N} \geq 0.05
g_*/\eta^2$, with $\eta\leq 1/(2 N_e)$, and so gives the lower bound ${\cal N}
\gtrsim 0.2g_*N_e^2$. {}For example for $g_*= 288.75$ and $N_e \simeq 50$ it
gives ${\cal N} \gtrsim 1.2\times 10^5$, which again is rather large.

\begin{figure}[t]
  \hfil\scalebox{0.5} {\includegraphics{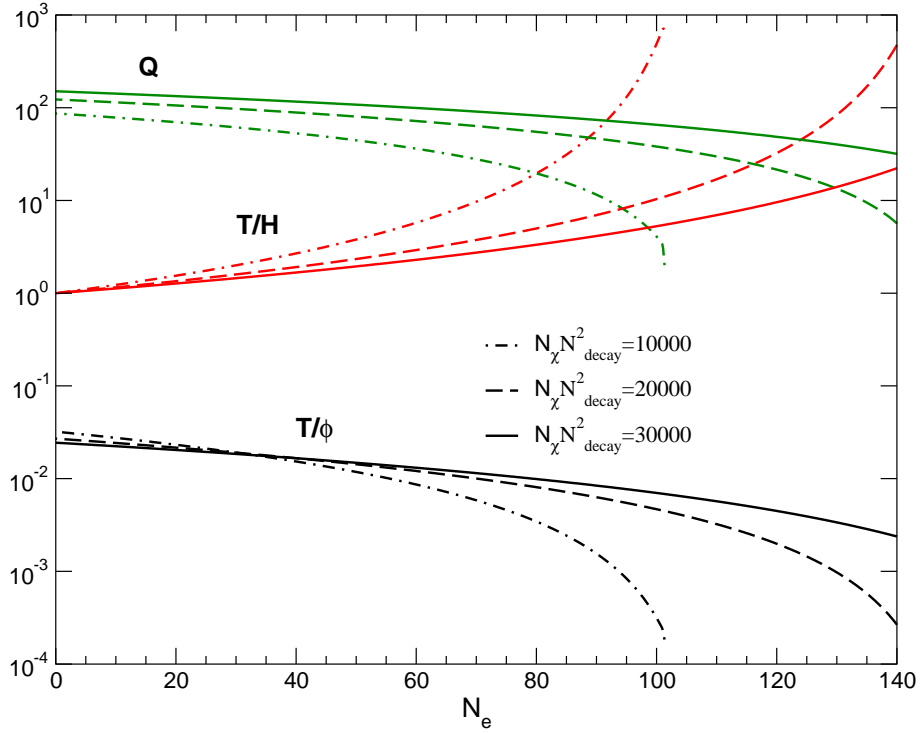}}\hfil
\caption{Warm  inflation  for hybrid  quadratic  potential:  Evolution of  the
  ratios $Q$ (top lines), $T/H$ (middle lines) and $T/\phi$ (bottom lines)
  depending on the number of e-folds to the end of inflation, for different
  values of ${\cal N}$=10000, 20000, 30000, with $g_*=228.75$,
  $\phi(0)/m_P=0.21$, $\eta =3$, and $V_0^{1/4}/m_P=3\times 10^{-4}$.}
\label{wihp}
\end{figure}

\vspace{3mm} {\bf Case $n=2$: Hybrid Quadratic potential.}  Figure
(\ref{wihp}) shows results for the evolution of the ratios $Q$, $T/H$ and
$T/\phi$, for ${\cal N} \equiv N_{\chi} N_{decay}^2 = 10000, 20000, 30000$ and
$g_*= 228.75$.  The end of inflation was taken as the point at which $T=H$,
and the analytical approximations do not any longer hold, and therefore in the
figure $N_e$ counts the number of e-folds left to the end of inflation.  The
condition $T>H$ in turn translates into a lower bound for ${\cal N}$:
\begin{equation}
{\cal  N}  \gtrsim  0.052  g_*  \left( \frac{Q_0}{\eta}  +  \frac  {2  N_e}{7}
\right)^2 \,. \label{Ngstarqd}
\end{equation}

\begin{figure}[t]
  \hfil\scalebox{0.5} {\includegraphics{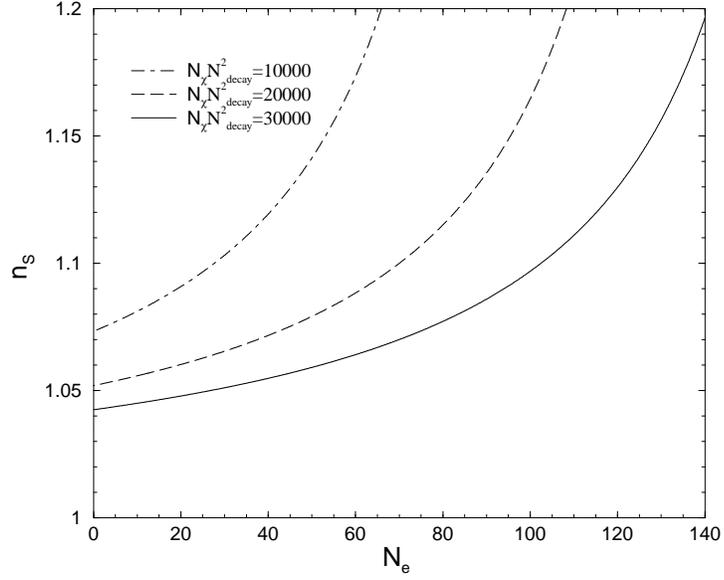}}\hfil
\caption{Hybrid quadratic  potential: Spectral index  depending on the  no. of
  e-folds to the end of inflation, for different values of ${\cal
    N}$=10000,20000,30000, with $g_*=228.75$, $\phi(0)/m_P=0.21$, $\eta=3$,
  and $V_0^{1/4}/m_P=3\times 10^{-4}$.}
\label{wihp2}
\end{figure}

{}From the approximated expression for the spectral index, Eq.
(\ref{nsapprox}), if one wants to keep $n_S$ within the observable range, it
requires $\eta/Q_0 \leq 0.093$, which for $N_e \simeq 50$ gives the lower
bound ${\cal N} \gtrsim 32.5 g_*$. Again, having slow-roll warm inflation for
example with $g_*\simeq 228.75$ needs ${\cal N} \gtrsim 7500$, but for $g_*
\simeq 10$ it only requires ${\cal N} \gtrsim 325$. As an example, {}Fig.
\ref{wihp2} shows the predicted spectral index depending on the number of
e-folds left to the end of inflation, for ${\cal N} \equiv N_{\chi}
N_{decay}^2 = 10000, 20000, 30000$ and $g_*= 228.75$. The corresponding
spectral index of the primordial spectrum would be that at around 50-55
e-folds, which is always $n_S < 1.2$. The value of the running can be obtained
from Eq.  (\ref{runapprox}), and it is given respectively by $n^\prime_S
\simeq -2.5\times 10^{-3},\; -8.5 \times 10^{-4},\; -4.7 \times 10^{-4}$.

Controlling the amplitude for the primordial spectrum from getting too large
was found to require values of $\eta/Q_0$ as large as possible, but not too
large values of $Q_0$. For values of ${\cal N}$, $g_*$ within the range of Eq.
(\ref{Ngstarqd}), the amplitude remains below say $10^{-4}$ for values of
$Q_0$ of order ${\cal O}(10)$.  Therefore in these kinds of models parameter
values can be found giving rise to the right order of magnitude for the
primordial spectrum in the strong dissipative regime, but the stronger
constraint comes from avoiding an overly blue-tilted spectrum.
The results for this example also reveal one generic feature of
warm inflation solutions in quantum field theory models,
that the total duration of inflation tends to be
small, of order the observational requirement of $\approx 60$ e-folds.
This is in contrast to cold inflation models, which in general cases can
predict huge numbers of e-folds, orders of magnitude larger
that the observational lower bound.  This fact about small
e-folds in warm inflation has been explored as a possible
solution to the low quadrapole observed in the CMB data
\cite{Berera:1997wz}.
Moreover, the small e-folds predicted in warm inflation could also
be a possible benefit to the transplankian problem,
in which small e-folds of inflation are 
preferred \cite{Martin:2000xs,Niemeyer:2000eh}.

\vspace{3mm} {\bf Case $n \geq 4$: Hybrid quartic and higher powers.}  In this
case having 50 e-folds in the strong regime requires for example ${\cal
  N}=N_\chi N_{decay}^2 \gtrsim 10^3$ for $g_* \simeq 10$, and ${\cal N}
\gtrsim 10^4$ for $g_* \simeq 228.75$. In addition if we want to get the right
amplitude for the spectrum and spectral index, 
it increases ${\cal N}$ by one
order of magnitude, since
the quartic coupling $a^2=\lambda$ has to be
adjusted to rather small values, and then the initial value of the field to
larger values to have $\phi(0)/T \geq 10$. Numbers do not change much whether
10 or 50 e-folds of inflation are demanded in the strong dissipative regime.

\subsection{hilltop potential}

\begin{figure}[htbp]
  \hspace{5cm}\epsfig{file=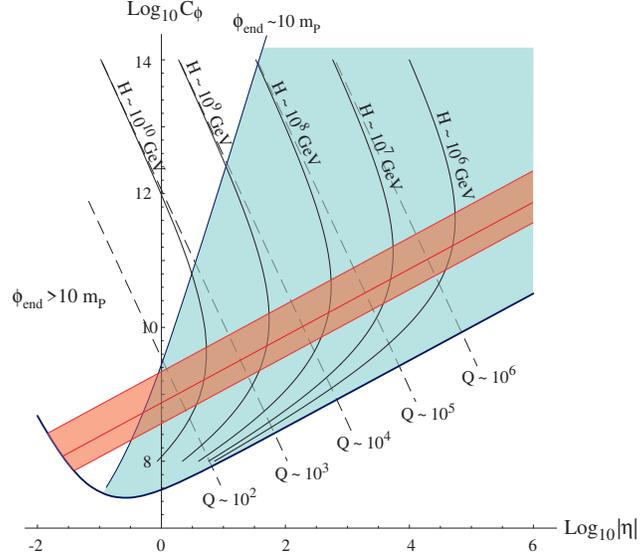, height=7.5cm}\caption{The figure
    shows the $\log C_\phi$-$\log |\eta|$ plane, where $C_{\phi} \equiv 0.64
    h^4 N_{\chi} N^2_{\rm decay}$.  The light grey (light green) area
    represents the space where the thermal spectrum of perturbations matches
    the observed amplitude $N_*\sim50$ $e$-foldings before the end of
    inflation, with \mbox{$\phi_{\rm end}<10\,m_P$}.  Within the dark grey
    (reddish) band, the resulting value of the spectral index is within the
    1-$\sigma$ window: \mbox{$n=0.960^{+0.014}_{-0.013}$}, as inferred by
    WMAP+BAO+SN data in the $\Lambda$CDM model for negligible tensor
    perturbations. In the graph lines of constant $H$ and
    \mbox{$Q\equiv\Upsilon/3H$} are also depicted.}
\label{figjuan}
\end{figure}

In \cite{BuenoSanchez:2008nc} the warm hilltop model \cite{Boubekeur:2005zm}
was investigated for the potential
\[\label{pot}
V=V_0-\frac{1}{2}|m^2|\phi^2+\ldots,
\]
where $V_0=3H^2m_P^2$, $m^2=V''(0)$, and the dots represent higher order terms
that become important only after relevant scales exit the horizon during
inflation.  The inflaton field $\phi$ is coupled to the fields of the two
stage mechanism in Eq.  (\ref{wi2stage}).  The scenario starts with the
inflaton field close to the hilltop. In \cite{BuenoSanchez:2008nc} this model
was constrained to obtain adequate e-folds of warm inflation and a consistent
amplitude for density perturbations.  In addition constraints were placed to
avoid gravitino overproduction.  The resulting parameter space for the strong
dissipative warm inflation regime is shown in {}Fig. \ref{figjuan}, where
$C_{\phi} \equiv 0.64 h^4 N_{\chi} N^2_{\rm decay}$. Moreover 
nongaussian effects in warm inflation 
\cite{Gupta:2002kn,Moss:2007cv,Chen:2007gd}
have been studied for this model in \cite{BuenoSanchez:2008nc}.  
In the strong
dissipative regime, there are in general large nongaussian effects
\cite{Moss:2007cv,Chen:2007gd}.
In particular it was shown in
\cite{Moss:2007cv} that entropy fluctuations during warm inflation play an
important role in generating non-Gaussianity, with the prediction
\begin{equation}
  -15  \ln \left(1+  \frac{Q}{14}\right) -  \frac{5}{2} \alt  f_{\rm  NL} \alt
  \frac{33}{2} \ln \left(1+ \frac{Q}{14}\right) - \frac{5}{2} \,,
\end{equation}
where $f_{\rm NL}$ is the non-linearity parameter and $r \equiv \Upsilon/3H$.
{}For the warm inflation results in Fig. \ref{figjuan}, $Q$ ranges from $10$
to $10^6$, and this implies from the above equation that $|f_{\rm NL}|$ ranges
from $10$ to $180$.
This is an interesting result in light of the recent WMAP analysis: the third
year CMB data \cite{Yadav:2007yy} gives $26.9 < f_{\rm NL} < 146.7$ at $95\%$
confidence level, although the five-year WMAP data \cite{Komatsu:2008hk} give
the limit $-9 < f_{\rm NL} < 111$ ($95\%$ CL).  The latest data show then a
tendency for $f_{NL}>0$, although this still will need to be confirmed by
future data, and in particular by data from Planck surveyor satellite
\cite{planck}.  If this is the case, this would disfavor conventional cold
inflation models which generally yield very low values of $f_{\rm NL} \lesssim
1$. On the other hand, the strong dissipative warm inflation regime, such as
the one found in this hilltop model, would be consistent with a non-gaussian
signal. For example, in this case taking $f_{\rm NL} \alt 110$ translates in
an upper limit on $Q \alt 1.3\times 10^4$, and therefore from
{}Fig.~\ref{figjuan} on a lower limit on the scale of inflation given by
\mbox{$H\gtrsim 10^8$ GeV}.

\section{Conclusions and Future Perspectives }
\label{concl}

Generically, the inflaton interacts with other fields in any typical inflation
model and so its dynamics is dissipative.  As such, inflation, like most
dynamics in nature is an open system phenomenon, thus requiring a much more
complex analysis of its dynamics than the one typically formulated for the
cold inflation picture.  This general point has been voiced by B. L. Hu and
coworkers ~\cite{hu1} and more specifically in the initial motivating papers
of warm inflation \cite{Berera:1995wh,Berera:1995ie,Berera:1996nv,BGR}.  In
all these cases, the point has been made that the problem of inflation
encompasses many different branches of Physics, from nonequilibrium
statistical dynamics to particle physics phenomenology.

Warm inflation dynamics is a rich area of study for both quantum field theory
real time dynamics and for particle physics model building.  The study of the
warm inflation dynamics has almost exclusively motivated the understanding of
strong dissipative behavior in quantum field theory.  This started initially
with the work of Berera, Gleiser and Ramos~\cite{BGR}, in which dissipation
was examined in the overdamped regime for the first time in quantum field
theory using extended linear response calculations.  Since then, specific
progress has been made to underpin interaction structures in interacting
quantum field theory models which lead to strong dissipative behavior under
warm inflationary conditions~\cite{BR1,BRplb1,BRfrw,BRplb2}.  At a more
general level, this work has motivated the first calculations of dissipation
in the low temperature regime~\cite{mx}.  In this review, a physical picture
has also been developed in Sec.~\ref{fd}, for explaining the dissipation
behavior found in warm inflation from the quantum field theory calculations,
and this is further developed in~\cite{grahammoss}.

Up to now all these studies has been based on variants of linear response
methods, in which case the dynamics is supposed to happen close to
equilibrium, or in a quasi-adiabatic regime, including various types of
resummations.  Several extensions to this work, which will give a more
accurate understanding of dissipation, are under way. {}For instance, when
departing from the quasi-adiabatic regime for the field dynamics, 
it is expected
that the local Markovian approximation commonly used to analyse the field
equations can differ significantly from the exact nonlocal equations.
Preliminary tests have shown that this difference can be very large for a
period of time starting from the initial period, but with 
differences between the
dynamics getting smaller at longer times, depending on the
model parameters~\cite{FRS}.

The methods used in this review are all based on the effective equation of
motion derived from the action functional.  In order to study strong
nonequilibrium dynamics, it requires methods beyond these.  In this case, full
kinetic set of equations for the relevant fields have to be studied, which
contain information not only of the inflaton effective dynamics but also about
thermalisation and equilibration. Work in this direction, also can help to
better understand the physics of particle production during the system
dynamics and be used to check the reliability of using the approximation of
thermal initial conditions for the bath fields.  Initial work in this
direction, though not directly related to the type of models relevant to warm
inflation as discussed extensively e.g. in Sec.~\ref{particlemod}, has
appeared~\cite{Aarts:2007ye}, while a work more on the warm inflation dynamics
motivated side is under way~\cite{BMR}.

This review has presented in detail the methods used so far to understand warm
inflation dynamics as well as the limitations of these studies.  In particular
these methods used so far are primarily quasi-adiabatic approximations for the
fields with the assumption of near thermal equilibrium evolution.  Despite
these limitations, these results find parameter regimes in which physically
acceptable solutions exist over time periods sufficiently
long to be of use in studies of warm inflation.
During this time interval in which these approximations
apply, and where a local Markovian dynamics can be used, in
contrast to the full non Markovian one, in the typical model implementations
discussed in Sec.~\ref{particlemod}, the amount of radiation production was
seen to be sufficient to change the cold inflationary picture predictions
regarding the density perturbations, thus requiring its description in terms
of the warm inflation picture. Moreover, once dissipative effects are
strong enough, inflation can be sustained and driven
longer than when these
effects are neglected.  Consequently, parameter values typically required in
the cold inflation case can be relaxed, which can help to evade
various problems that plaque the standard scenario of inflation, like the
graceful exit and $\eta$-problem, discussed in details in this review, as
well as the problems of
quantum-to-classical transition~\cite{Berera:1995ie,bel}
and the initial conditions for inflation~\cite{BG,ROR}.

The development of the quantum field theory dynamics of warm inflation has in
turn been applied to particle physics model building, in which warm inflation
dynamics in realised.  Early on it was recognisied in \cite{Berera:1999ws} that
warm inflation has some appealing and unique model building features.  In
particular, it offers a simple solution to the $\eta$-problem and for
monomial potentials, observationally consistent inflation can occur for the
inflaton amplitude below the Planck scale $\langle \phi \rangle < m_P$.  These
features have been realised in explicit first principles quantum field theory
models of warm inflation in \cite{BGR2,Berera:1999ws,bb4,BuenoSanchez:2008nc}.
These studies are now being extended to develop a complete particle cosmology
in which not only is warm inflation realised, but in addition other features
such as leptogenesis and gravitino abundances are addressed,
developing in depth some of the work already started on
these topics \cite{Taylor:2000jw,bb1,bb4,Lambiase:2006md,BuenoSanchez:2008nc}.  In a separate
direction, the warm inflation models developed so far in
\cite{bb4,BuenoSanchez:2008nc} have been in the low temperature regime.  In
\cite{bbrecent} this is being extended to higher temperatures.  This requires
calculating all the thermal loop corrections in the SUSY models that have the
two-stage interaction structure Eq. (\ref{wi2stage}) relevant for warm
inflation.  Some initial work has been done in \cite{Hall:2004zr}, and a more
detailed analysis is now underway \cite{bbrecent}.

\acknowledgments

This work was partially supported by a U.K. Particle Physics and Astronomy
Research Council (PPARC) Visiting Researcher Grant.  In addition A.B.  is
partially supported by STFC and R.O.R. is partially supported by Conselho
Nacional de Desenvolvimento Cient\'{\i}fico e Tecnol\'ogico (CNPq - Brasil)
and Funda\c{c}\~ao de Amparo \`a Pesquisa do Estado do Rio de Janeiro
(FAPERJ).

\end{document}